\documentclass[a4paper,11pt]{article}
\usepackage{natbib}
\usepackage{xr}
\usepackage{xr-hyper}
\usepackage[hidelinks]{hyperref}
\usepackage[singlespacing]{setspace}
\usepackage{amsmath}
\usepackage{booktabs}
\usepackage{longtable}
\usepackage{array}
\usepackage{multirow}
\usepackage{colortbl}
\usepackage{pdflscape}
\usepackage{xcolor}
\usepackage{url}
\usepackage{array,dcolumn}
\usepackage{caption}
\usepackage{graphicx}
\usepackage{appendix}
\usepackage{subcaption}

\newcolumntype{d}[1]{D{.}{.}{#1} }
\usepackage[flushleft]{threeparttable}
\usepackage[capposition=top]{floatrow}

\usepackage{authblk}

\title{\flushleft The Value of Research Funding for Knowledge Creation and Dissemination:  A study of SNSF Research Grants}
\author[1]{Rachel Heyard}
\author[2]{Hanna Hottenrott}
\affil[1]{\small Swiss National Science Foundation, Bern, Switzerland (rachel.heyard@snf.ch)}
\affil[2]{\small TUM School of Management, Munich, Germany (hanna.hottenrott@tum.de)}
\date{}

\begin{document}

\maketitle

\begin{abstract}
This study investigates the effect of competitive project-funding on researchers' publication outputs. Using detailed information on applicants at the Swiss National Science Foundation  and their proposal evaluations, we employ a case-control design that accounts for individual heterogeneity of researchers and selection into treatment (\textit{e.g.} funding). We estimate the impact of grant award on a set of output indicators measuring the creation of new research results (the number of peer-reviewed articles), its relevance (number of citations and relative citation ratios), as well as its accessibility and dissemination as measured by the publication of preprints and by altmetrics. 
The results show that the funding program facilitates the publication and dissemination of additional research amounting to about one additional article in each of the three years following the funding. The higher citation metrics and altmetrics by funded researchers suggest that impact goes beyond quantity and that funding fosters dissemination and quality. 

\end{abstract}

\section*{Keywords}
Research funding, altmetrics, scientific articles, preprints, research productivity, impact evaluation

\clearpage

\hypertarget{introduction}{%
\section{Introduction}\label{introduction}}

Scientific research generated at universities and research organisations plays an important role in knowledge-based societies \citep{Fleming2019,Poegeetal2019}. The created knowledge drives scientific and technological progress and spills over to the broader economy and society \citep{Jaffe1989,Stephan_2012,Hausman_2020}. The growing importance of science-based industries puts additional emphasis on the question how scientific knowledge is generated and whether public funding can accelerate knowledge creation and its diffusion. 
In an effort to promote scientific research, grant competitions as a means of allocating public research funding have become an important policy tool \citep{Froumin2015,Oancea2019}. The goal is to incentivize the generation of ideas and to allocate funding such that it is most likely to deliver scientific progress and eventually economic and social returns\footnote{The importance of competitive research funding increased substantially over the past three decades. The basic idea of promoting such science policy goes back to New Public Management reforms which aimed to increase the returns to public science funding through the selective provision of more funding to the most able researchers, groups and universities (winners in funding competitions), and to create performance incentives at all levels of the university system \citep{KruckenMeier_2006,GlaserVelarde_2018}.}. In light of these developments, it is important to understand whether research grants indeed facilitate additional, relevant research outputs and whether these are accessible to the public.

In particular individual-level analyses are highly interesting since most grants are awarded to individual researchers or to small teams of researchers. The estimation of the effect that a grant has on research outputs is, however, challenging. The  main difficulties are the availability of information on all applicants (not only winners) as well as detailed information about the individual researchers (demographic information). Moreover, the non-randomness of the award of a grant through the selection of the most able researchers into the funding program results in the non-comparability of funded and non-funded researchers. The fact that researchers can receive multiple grants at the same time as well as several consecutive grants further challenges the estimation of effects from funding \citep{Jaffe_2002}. Another difficulty stems from finding appropriate measures for research output \citep{Oancea2019}. Publications and citations are easy to count, but likely draw an incomplete picture of research impact, its dissemination and the extent to which funded research contributes to public debates. Moreover, both publication and citation patterns as well as funding requirements are highly field-dependent which makes output analyses in mixed samples or inter-disciplinary programs difficult.  \\
In this study, we aim to quantify the effect of the Swiss National Science Foundation's (SNSF)\footnote{The SNSF is Switzerland's main research funding agency. The SNSF is mandated by the Swiss confederation to allocate research funding to eligible researchers at universities, (technical) colleges and research organisations.} project funding (PF) grants on the individual researcher in terms of future scientific publications and their dissemination. Our analyses is based on detailed information on both grants and awardees covering 20'476 research project grants submitted during the period 2005 and 2019.
This study adds to previous work in several dimensions. By focusing on the population of applicants which constitutes a more homogeneous set of researchers than when comparing grant winners to non-applicants and by accounting of individual characteristics of the applicants, our study results are less prone to overlook confounding factors affecting both the likelihood to win a grant as well as research outputs. Information on the evaluation scores submitted in the peer-review process of the grant proposals allows us to compare researchers with similarly rated proposals. In other words, by comparing winning applicants to non-winners and by taking into account the evaluation scores that their applications receive, we can estimate the causal effect of the grant on output while considering that both research ideas as well as grant writing efforts (and skills) are required for winning a grant. By studying a long time-period and accounting for the timing of research grants and outcomes, we can further take into account that there are learning effects from the grant writing itself even for unsuccessful applicants \citep{Ayoubietal_2019}. To benchmark our results to previous studies, we first investigate the impact of grants on publication outputs. In addition, we consider preprints which have become an important mode of disseminating research results quickly, but received so far no attention in research of funding effects. Preprints do not undergo peer-review \citep{Bergetal2016,Serghiou2018}, but help researchers to communicate their results to their community and to secure priority of discovery.

This study goes beyond previous work that mainly considered citation-weighted publication counts, by measuring impact in a researcher's field of study by relative citation ratios (RCR) and field citation ratios (FCR). These metrics account for field-specific citation patterns. Additionally we explicitly explore researchers' altmetric scores as a measure of attention, research visibility and accessibility of research outcomes beyond academia. Altmetrics reflect  media coverage, citations on Wikipedia and in public policy documents, on research blogs and in bookmarks of reference managers like Mendeley, as well as mentions on social networks such as Twitter. While altmetrics may reflect fashionable or provocative research, they may indicate accessible insights disseminated through the increasingly important online discussion of research and may therefore measure the general outreach of research  \citep{Warrenetal2017}. 
Although they are a potentially important measure of dissemination to the wider public and therefore of research impact in the age of digital communication \citep{BORNMANN2014,Konkiel2016,Lazaroiu2017}, the effect of funding on altmetrics has not been investigated so far.

Finally, by explicitly investigating outputs over several years after funding, our study contributes new insights on the persistency of effects. Since a large share of project funding typically goes into wages of doctoral and post-doctoral researchers which require training and learning on the job, there may be a considerable time-lag between the start of the project and the publication of any research results and an underestimation of output effects when considering only immediate outcomes.

The results from our analysis based on different estimation methods show that grant-winning researchers publish about one additional peer-reviewed publication more per year in the three years following funding than comparable but unsuccessful applicants. Moreover, these publications are also influential as measured by the number of citations that they receive later on.
SNSF PF seems to promote timely dissemination as indicated by the higher number of published preprints and researchers' higher altmetrics scores. The funding impact is particularly high for young(er) researchers as well as for researchers at a very late career stage when funding keeps output levels high. 
These results add new insights to the international study of funding effects which provided partially ambiguous findings as our review in the next section illustrates. In summary, the results presented in the following stress the important role played by project funding for research outcomes and hence for scientific progress. Institutional funding alone does not appear to facilitate successful research to the same extent as targeted grants which complement institutional core funds. 

\subsection{The impact of funding on research outcomes}
The impact of competitive research funding on knowledge generation (typically proxied by scientific publications) has been studied in different contexts and at multiple levels: the institutional level, the research group or laboratory, and the level of the individual researcher.   
At the level of the university, \citet{AdamsGriliches_1998} find a positive elasticity of scientific publications to university funding. \cite{Payne_2002} and \cite{PayneSiow_2003}, using congressional earmarks and appropriation committees as instruments for research funding, present similar results. They show that a \$1 million increase in funding yields 10 to 16 additional scientific articles. \cite{Wahls_2018} analyses the impact of project grants from the National Institutes of Health (NIH) in the United States and finds positive institution-level returns (in terms of publications and citation) to funding which, however, diminish at higher levels of funding. 

At the laboratory level, the results are rather inconclusive so far which is likely due to heterogeneity in unobserved lab characteristics and the variety of grants and resources that typically fund lab-level research. An analysis of an Italian biotechnology funding program by \cite{Aroraetal_1998} finds a positive average elasticity of research output to funding, but with a stronger impact on the highest quality research groups. These findings, however, seem to be specific to engineering and biotechnology. \citet{CarayolMatt_2004} included a broader set of fields and did not find a strong link between competitive research funding and lab-level outputs.  

At the level of the individual researcher, \cite{AroraGambardella_2005} find that research funding from the United States National Science Foundation (NSF) in the field of Economics has a positive effect on publication outcomes (in terms of publication success in highly ranked journals) for younger researchers. For more advanced principle investigators (PIs between 5 and 15 years since PhD), however, they do not find a significant effect of NSF funding when taking the project evaluation 
into account. \cite{JacobLefgren_2011} study personal research funding from the NIH and find that grants resulted in about one additional publication over the next five years. These results are close to the estimated effect from public grants of about one additional publications in a fixed post-grant window in a sample of Engineering professors in Germany \citep{HottenrottThorwarth2011}. Likewise, a study on Canadian researchers in nanotechnology \citep{BEAUDRY2012} documents a significant positive relationship between public grants and the number of subsequently published articles. 

More recent studies, considered output effects both in terms of quantity and quality or impact. Evaluating the impact of funding by the Chilean National Science and Technology Research Fund on research outputs by the PIs, \cite{BENAVENTE2012} find a positive impact in terms of number of publications of about two additional publications, but no impact in terms of citations to these publications.
In contrast to this, \cite{Beaudry2019} show that there is also an influence of public grants (unlike for private sector funding) on the number of citations for nanotechnology researchers in Canada. In addition, \cite{HottenrottLawson_2017} find that grants from public research funders in the United Kingdom contribute to publication numbers (about one additional publication per year) as well as to research impact (measured by citations to these publications)
even when grants from other private sector sources are accounted for. Results for a sample of Slovenian researchers analysed by \cite{Malietal2017}, however, suggest that public grants result in `excellent publications'\footnote{Excellent publications in this study were for instance papers in the upper quarter of journals included in the Science Citation Index (SCI).} only if researchers' funding comes mostly from one source. 

Explicitly looking at research novelty\footnote{Novelty is measured by the extent to which a published paper makes first time ever combinations of referenced journals while taking into account the difficulty of making such combinations.}, \citet{WANG2018} find that projects funded by competitive funds in Japan have on average higher novelty than projects funded through institutional funding. 
However, this only holds for senior and male researchers. For junior female researchers, competitive project funding has a negative relation to novelty. 

In a study on Switzerland-based researchers, \citet{Ayoubietal_2019} find, in a sample of 775 grant applications for special collaborative, multi-disciplinary and long-term projects, that participating in the funding competition does indeed foster collaborative research with co-applicants. For grant-winners, they observe a lower average number of citations received per paper compared to non-winners (not controlling for other sources of funding that the non-winners receive). The authors relate this finding to the complexity of such interdisciplinary projects, the cost of collaboration, and the fact that also applicants who do not eventually win this particular type of grant publish more as a result of learning from grant writing or through funding obtained from alternative sources.\\

By studying grants distributed via the main Swiss research funding agency, we are capturing the vast majority of competitive research grants in the country. The Swiss research funding system is characterised by a relatively strong centralisation of research funding distribution with the SNSF accounting by far for the largest share of external research funding of universities \citep{Schmidt2008,Jonkers2016}\footnote{Charities and private sector grants do play an increasing, but still relatively minor role in Switzerland \citep{Schmidt2008, Jonkers2016}.}. 
To account for major sources outside of Switzerland such as from the European Research Council (ERC), we collected information on Swiss-based researchers who received such funding during our period of analysis.

\section{Empirical model of funding and research outputs}\label{sec:empirical-model}

All of the following is based on the assumption that academic researchers strive to make tangible contributions to their fields of research. The motivations for doing so can be diverse and heterogeneous ranging from career incentives to peer-recognition \citep{Franzoni2011}. We also assume that producing these outputs requires resources (personnel, materials, equipment) and hence researchers have incentives to apply for grants to fund their research. However, research output, that is the success of a researcher in producing results and the frequency with which this happens, also depends on researcher characteristics, characteristics of the research field and the home institution. Research success is also typically path-dependent following a success-breeds-success pattern. Thus, we build on the assumption that a researcher who generates an idea for a research project files a grant application to obtain funding to pursue the project. If the application succeeds, the researcher will spend the grant money and may or may not produce research outputs. The uncertainty is inherent to the research process. The funding agency screens funding proposals and commissions expert-reviews to assess the funding worthiness of the application. If the submitted proposal received an evaluation that is sufficiently good in comparison to the other proposals, funding is granted in accordance with the available funding amount. 
This implies that even in case of a rejected grant proposal the researcher may pursue the project idea, but without these dedicated resources available. In many instances funding decisions are made at the margin, with some winning projects being only marginally better than non-wining projects \citep{Gravesd2011,Neufeld2013,Fang2016}. If the funding itself has indeed an effect on research outcomes, we would expect that the funded researcher is more successful in generating outputs both in terms of quantity and quality. 

In addition to resource-driven effects, there may also be direct dissemination incentives related to public project funding. On the one hand, funding agencies may encourage or even require dissemination of any results from the funded project. On the other hand, the researchers may have incentives to publish research outcomes to signal project success to the funding agency and win reputation gains valuable for future proposal assessments. 

While estimating the contribution of funding to research outputs measured by different indicators, we have to take into consideration that the estimation of the funding effect requires assumptions about output generation by researchers. The extent to which the output produced can be attributed to the funding itself also depends on the econometric model used \citep{Silberzahn2018}. We therefore take a quantitative multi-method approach taking up and adding to methods applied in previous related studies. Comparing the results from different estimation methods also allows an assessment of the sensitivity of our conclusions to specific modelling assumptions.  In particular, we estimate longitudinal regression models which aim to account for unobserved heterogeneity between researchers. In addition, we use non-parametric matching methods to explicitly model the selectivity in the grant awarding process.   

\subsection{Mixed Effects Models}
We define \(P_{it}\) as the research output of researcher \(i\) in year \(t\) and \(F_{it-1}\) as a binary variable indicating whether this same researcher \(i\) had access to SNSF funding in year \(t-1\). Note that this indicator takes the value one for the entire duration of the granted project. The funding information is lagged by one year as an immediate effect of funding on output is unlikely. Note that, we will differentiate between funding as PI and as co-PI (only).
The general empirical model can then be expressed as

\[P_{it}(\phi) = \phi \ [F_{it-1} + X_{it} + T_t] + v_{i} + \epsilon_{it},\]

with \(\phi\) being the vector of parameters. \(X_{it}\) represents a vector with explanatory factors at \(t\) including observed characteristics of the researcher and the average quality of the grant applications as reflected in the average evaluation score. Further \(T_t\) captures the overall time trend, \(v_i\) is the unobserved individual heterogeneity, and \(\epsilon_{it}\) is the error term.

The specification above describes a production function for discrete outcome following \citet{Blundell}. 
As a first estimation strategy, count data models will be used to estimate research outputs, as for example the number of peer reviewed articles or preprints. 
Moreover, these models account for unobserved individual characteristics, \(v_i\), which likely predict research outputs besides observable characteristics and are independent from project funding. One way to estimate this unobserved heterogeneity is to use random intercepts for the individuals\footnote{An alternative approach is to employ pre-sample information of the researcher as a proxy for unobservable characteristics, such as a researcher's ability or writing talent which impact research output in the (later) sample period. We conducted such linear feedback models (LFM) as robustness test and present them in Supplement \ref{supp-analyses}}, here the researchers, and account for the hierarchical structure of the information (\emph{e.g.} panel data). Thus, we estimate mixed count models to capture \(v_i\)\footnote{We use the \texttt{lmer} package in \texttt{R} and a negative binomial family.}. The mixed regression models for count data take the following form
\[\log \mbox{E} (P_{it} \ | \ \mbox{data})  =  \phi \ [F_{it-1} + X_{it} + T_t]+ v_{i} \ .\]

In addition to count-type outputs, we estimate the effect of funding on continuous output variables such as the average number of yearly citations per article or the researcher's average yearly altmetric score. For these output types we estimate linear regression models based on a comparable model specification with regard to \(F_{it-1}, X_{it}, T_t\) and \(v_{i}\). \\

\subsection{Non-Parametric Treatment Effect Estimation}\label{subsec:empirical-model}
In an alternative estimation approach, we apply a non-parametric technique: The average treatment effect of project funding on scientific outcomes is estimated by an econometric matching estimator which addresses the question of “How much would a funded researcher have published (or how much attention in terms of altmetrics or citations would her research have received) if she had not received the grant?”. This implies comparing the actually observed outcomes to the counterfactual ones to derive an estimate for the funding effect. Given that the counterfactual situation is not observable, it has to be estimated. 

For doing so, we employ a nearest neighbor propensity score matching. That is, we pair each grant recipient with a non-recipient by choosing the nearest `twin' based on the similarity in the estimated probability of receiving a grant and the average score that the submitted applications received. Note that we select the twin researcher from the sample of unsuccessful applicants so that matching on both, the general propensity to win (which includes personal and institutional characteristics) and the proposal's evaluation score, allows to match both on individual as well as on proposal (or project idea) characteristics to find the most comparable individuals. 

The estimated propensity to win a grant is obtained from a probit estimation on a binary treatment indicator which takes the value of one for each researcher-year combination in which an individual had received project funding. The advantage of propensity score matching compared to exact matching is that it allows to combine a larger set of characteristics into a single indicator avoiding the curse of dimensionality. Nevertheless, introducing exact matching for some key indicators can improve the balancing of the control variables after matching. In particular, we match exactly on the year of the funding round as this allows to have the same post-treatment time window for treated and control individual and also
captures time trends in outputs which could affect the estimated treatment effect. In addition, we match only within a research field to not confound the treatment effect with heterogeneity in resource requirements and discipline differences in output patterns.
We follow a matching protocol as suggested by \cite{GerfinLechner2002} and calculate the Mahalanobis distance between a treatment and a control observation as \[MD_{ij} = (Z_i-Z_j) \Omega^{-1} (Z_i-Z_j) \] 
where $\Omega$ is the empirical covariance matrix of the matching arguments (propensity score and evaluation score). We employ a caliper to avoid bad matches by imposing a threshold of the maximum distance allowed between the treated and the control group. That is, a match for researcher $i$ is only chosen if $|Zj - Zi| < \epsilon$, where $\epsilon$ is a pre-specified tolerance.
After having paired each researcher with the most similar non-treated one, any remaining differences in observed outcomes can be attributed to the funding effect. The resulting estimate of the treatment effect is unbiased under the conditional independence assumption \citep{Rubin1977}. In other words, in order to overcome the selection problem, participation and potential outcome have to be independent for individuals with the same set of characteristics $X_{it}$\footnote{In addition to the closeness on $MD$, we use elements of exact matching by requiring that selected control researchers belong exactly to the same subject field and to be observed in the same year as the researchers in the treatment group. This allows to account for different publication patterns across disciplines and also for time trends in funding likelihood and in the outcome variables.}.
Note that by matching on the evaluation score in addition to the propensity score, our approach is similar to the idea of regression discontinuity design (RDD). The advantage of the selected approach is, however, that it allows us to draw causal conclusions for a more representative set of individuals. While RDD designs have the advantage of high internal consistency, this comes at the price of deriving effects estimates only for researchers around the cut-off \citep{delaCuesta2016}. Yet, in our case, this threshold is not constant, but depends on the pool of submitted proposals and there is considerable variation in the evaluation scores that winning proposals receive. In our application, we also expect heterogeneous impacts across researchers so that a local effect might be very different from the effect for researchers away from the threshold for selection \citep{BATTISTIN2008715}. 

Using the matched comparison group, the average effect on the treated can thus be calculated as the mean difference of the matched samples:	
\[\hat{\alpha}_{TT} = \frac{1}{n^t} \Biggl( \sum_{i}P_{i}{^T} - \sum_{j}\hat{P}_{j}{^C} \Biggr) \]
with $P_i^T$ being the outcome variable in the treated group, $P_j^C$ being the counterfactual for $i$ and $n^T$ is the sample size (of treated researchers).\footnote{As we perform sampling with replacement to estimate the counterfactual situation, an ordinary t-statistic on mean differences after matching is biased, because it does not take the appearance of repeated observations into account. Therefore, we have to correct the standard errors in order to draw conclusions on statistical inference, following \citet{Lechner2001}.}

\hypertarget{data-and-descriptive-analysis}{%
\section{Data and descriptive analysis}\label{data-and-descriptive-analysis}}

Data provided by the SNSF has been used to retrieve a set of researchers of interest. These researchers have applied to the SNSF funding instrument project funding (PF) or Sinergia \footnote{The Sinergia scheme is closely linked to PF, so that we will not differentiate between them in the following.}
as main applicant (\textit{e.g.} PI) or co-applicant\footnote{If granted, a co-applicant is entitled to parts of the funding.} (\textit{e.g.} co-PI). The PF scheme is a bottom-up approach as it funds costs of research projects with a topic of the applicant's own choice.  

The study period is dynamic and researcher-specific: it starts with the year in which the SNSF observes the researcher for the first time as (co-)PI to PF or as a career funding grantholder (after the postdoctoral level); the year the independent research career starts. However, this study period has its lower bound in 2005. The period ends in 2019 for everyone, and some researchers are observed for a longer period than others. For each researcher, a pre-sample period is defined, including the five years before the observation started. Pre-sample information on all outcome variables of interest is needed to account for heterogeneity between the individuals in the way that they enter the study in linear feedback models and for matching on ex-ante performance in the non-parametric estimation approach. Further, only researchers who applied at least once after 2010 to the SNSF are included to ensure a minimum research activity. 
In a next step, we retrieve a unique Dimensions-identifier (Dim-ID) from the \emph{Dimensions} database \citep{Dimensions} using a person's name, research field, age and information about past and current affiliations\footnote{If Dimensions found more than one ID for a certain name, we used further information on the researcher available to the SNSF to narrow the ID-options down. This supplementary information was, if present the ORCID, the current and previous research institution(s), country and birth year. Only researchers with a unique ID could be used in the following. See Table \ref{tab:percentages_found_not_found} for a comparison of the researchers that were found and not found}.
The Dim-ID enables us to collect disambiguated publication information for these researchers to be used in the empirical analysis. 

\hypertarget{variables-and-descriptive-statistics}{%
\subsection{Variables and descriptive statistics}\label{variables-and-descriptive-statistics}}

The original data set comprised 11'228 eligible researchers. 10\% (1'143) of the latter could not be identified in the Dimensions database. Among the researchers found using their name, the supplementary information from the SNSF database (country, ORCID, institution, etc.) did not match in 1\% of the cases, and we were not sure that we found the correct researcher. For 12\% of the researchers found in Dimensions no unique ID could be retrieved. After removing these observations, we observe a total of 8'793 distinct researchers (78\% of the eligible researchers\footnote{Some characteristics on the researchers without unique ID can be found in Table \ref{tab:percentages_found_not_found}}) and the final data set is composed of 82'249 researcher-year observations.  
On average researchers are observed for 9.35 years. The maximum observation length, from 2005 to 2019 is 15 years, and 2'319 researchers are observed over this maximal study period. All the publication data was retrieved in September 2020.

\hypertarget{research-funding}{%
\subsubsection{Research funding}\label{research-funding}}
The central interest of the study is the effect 
competitive project funding has on a researcher's subsequent research outputs.
The information on SNSF funding indicates whether a researcher had access to SNSF funding as a PI and/or co-PI in a certain year. We differentiate between PIs and Co-PIs to test whether the funding effect differs depending on the role in the project. On average the researchers in our data set are funded by the SNSF for 4.6 years during the observation period; for 3.3 years as PI of a project, see Table \ref{tab:output_baseline}. In total 20'476 distinct project applications (not necessarily funded) are included in the data. On average a PIs is involved in a total of 3.7 project applications (as PI or co-PI); in 3.1 submissions as PI, and in 2.3 submissions as co-PI. About 66\% of all projects in the data have one sole PI applying for funding, 22\% have a PI and a co-PI, 8\% a PI and two co-PIs, and 4\% are submitted by a PI together with three or more co-PIs. Note that the percentage of successful applications in our data set is 48\% over the whole study period (the success rate for the STEM applications is $\sim 60\%$, it is $\sim 44\%$ in SSH and the one of the LS is the lowest with $\sim 40\%$).

These numbers reflect that in the Swiss research funding system, project funding does play an important role, but that institutional core funding is also relatively generous. The latter accounts for - on average - more than 70\% of overall university funding \citep{Schmidt2008,Reale2017}. This allows researchers to sustain in the system without project funding. While overall, institutional funding is quite homogeneous across similar research organisations in the country, it differs between institution types. It is therefore important to account for institutional funding in the following analyses as it provides important complementary resources to researchers \citep{Jonkers2016}.
Moreover, within the different institution types, we also account for the research field and the career stage of researchers as this may also capture individual differences in core budgets. We present sample characteristics in terms of these variables in subsection  \ref{confounding-variables}.
Another important aspect to consider when analysing the effect of research funding is funding from other sources, other than institutional funding \citep{HottenrottLawson_2017}. In all European countries the European Research Council (ERC) plays an important role. Hence, we collected data on Swiss-based researchers who received ERC funding and matched them to our sample. Of all the researchers considered in this study only a small fraction (4.2\%) ever received funding by the ERC. Most of these researchers had a PF grant running at the same time (87\%). Figure \ref{fig:erc_evolution} shows the count of observations in the different funding groups in more detail.\footnote{Since only a few cases are identified to hold major international grants but no SNSF funding, we do not differentiate between these groups in the following. Note that the data was retrieved from the ERC Funded Projects Database included only grants acquired since 2007.}

\hypertarget{research-outputs-and-research-funding}{%
\subsubsection{Research outputs}\label{research-outputs-and-research-funding}}
Table \ref{tab:output_baseline} summarises the output measures as well as the funding length. The most straightforward research output measure is the number of (peer-reviewed) articles. On average, a researcher in our data publishes 4.9 articles each year. The annual number of articles is higher in the STEM (5.7) and life sciences (LS) (6.5) than in the Social Science and Humanities (SSH) where researchers published about 1.5 publications per year, on average. See Table \ref{tab:output_baseline_RA} for differences in all output variables (as well as funding and researcher information) by field. In some disciplines, such as biomedical research, physics, or economics, preprints of articles are widely used and accepted \citep{Bergetal2016,Serghiou2018}. As preliminary outputs they are made available early and thus are an interesting additional output, potentially indicating the dissemination and accessibility of research results. The average of the yearly number of preprints is a lot lower than the one of articles (0.4) which is due to preprints being a research output that emerged only rather recently and are more common in STEM fields than in others (see Table \ref{tab:output_baseline_RA}). Another output measure is the number of yearly citations per researcher. This is the sum of all citations of work by a certain researcher during a specific year to all her peer-reviewed articles published since the start of the observation period. Citations to articles published before the start of the observation period are not taken into account. On average a researcher's work in the study period is cited 132.9 times per year. This variable is however substantially skewed with 6.8\% of researchers accounting for 50\% of all citations and highly correlated with the overall number of articles that a researcher published. There are also field differences with the average citation numbers between the Life Sciences (185.2) and the STEM fields (157.7), but both numbers are substantially higher than in the SSH (25.6). The average number of citations per (peer-reviewed) article of a researcher is informative about the average relevance of a researcher's article portfolio. The articles in our sample are cited on average 4.2 times per year.

The \emph{altmetric score} of each article is retrieved as an attention or accessibility measure of published research. Following the recommendation by \cite{Konkiel2016}, we employ a `baskets of metrics' rather than single components of the altmetric score. 
This score is a product of Digital Science and represents a weighted count of the amount of attention that is picked up for a certain research output 
\footnote{\url{https://help.altmetric.com/support/solutions/articles/6000233311-how-is-the-altmetric-attention-score-calculated-}}. 
Note that the average altmetric score for a researcher at \(t\) is the mean of the altmetrics of all articles published in the year \(t\).\footnote{Unfortunately the altmetric cannot be retrieved as a time-dependent variable from Dimensions but only as the altmetric state at the time point of data retrieval (September 2020). Therefore the altmetric informs us on the cumulative importance an article published at \(t\) got until September 2020.} On average a researcher in our sample achieves an altmetric of 13. Similar to citation counts, this variable is heavily skewed. The differences in altmetrics across disciplines are rather small (see Table \ref{tab:output_baseline_RA}). \\
When using simple output metrics like citation counts, it is important to account for field-specific citation patterns. In order to do so, we collect the \textit{relative citation ratio} (RCR) and the \textit{field citation ratio} (FCR). The RCR was developed by the NIH \citep{Hutchins_2016}. As described by \cite{Surkis_2018}, the RCR uses an approach to evaluate an article's citation counts normalized to the citations received by NIH-funded publications in the same area of research and year. The calculation of the RCR implies to dynamically determine the field of an article based on its co-citation network, that is, all articles that have been cited by articles citing the target article. 
The advantage of the RCR is to field- and time-normalize the number of citations that an article received. A paper that is cited exactly as often as one would expect based on the NIH-norm receives an RCR of 1 and an RCR larger one indicated that an article is cited more than its expectation given the field and year. 
The RCR is only calculated for the articles that are present on PubMed, have at least one citation and are older than two years. Thus, when analysing this output metric, we focus on researchers in the life sciences only. The FCR is calculated by dividing the number of citations a paper has received by the average number received by publications published in the same year and in the same fields of research (FoR) category. Obviously, the FCR is very dependent on the definition of the FoR. Dimensions uses FoR that are closest to the Australian and New Zealand Standard Research Classification \citep{ANZSRC}. For the calculation of the FCR a paper has to be older than two years. Simlar to the RCR, the FCR is normalized to one and an article with zero citations has an FCR of zero. As the altmetric, the RCR and FCR cannot be retrieved time-dependently but are snapshots at the day of retrieval. We will refer to the average FCR/RCR at $t$, as the average of the FCRs/RCRs of the papers published in $t$. According to \cite{Hutchins_2016}, articles in high-profile journals have average RCRs of approximately 3. 
The key difference between the RCR and the FCR is that the FCR uses fixed definition of the research field, while for the RCR a field is relative to each publication considered. Table \ref{tab:output_baseline_RA} shows that the average rates are comparable across fields.

\begin{table}
\caption{Descriptive statistics for the output measures, funding measures and researcher characteristics.}\label{tab:output_baseline}
\centering \small
\begin{tabular}{llrrrl}
\toprule
  & \% (Mean) & SD & Min & Max & NAs\\
\midrule
\addlinespace[0.3em]
\multicolumn{6}{l}{\textbf{Output Measures}}\\
\hspace{1em}\# of articles & 4.9 & 7.2 & 0 & 222 & 0\\
\hspace{1em}\# of preprints & 0.4 & 1.5 & 0 & 54 & 0\\
\hspace{1em}\# of citations  & 132.9 & 321.1 & 0 & 7'888 & 0\\
\hspace{1em}\# of av. citations  & 4.2 & 4.9 & 0 & 146.2 & 0\\
\hspace{1em}Yearly av. altmetric  & 13.2 & 44.6 & 1 & 4'211 & 35'237 \\
\hspace{1em}Yearly av. FCR & 6.6 & 12.4 & 0 & 786.5 & 26'345 \\
\hspace{1em}Yearly av. RCR  & 1.6 & 3.6 & 0 & 242.2 & 42'352 \\
\rowcolor{gray!11}
\addlinespace[0.3em]
\multicolumn{6}{l}{\textbf{Funding Information}}\\
\rowcolor{gray!11}\hspace{1em}\# of years funded & 4.6 & 4.7 & 0 & 15 & 0\\
\rowcolor{gray!11}
\hspace{1em}\# of years funded as PI & 3.3 & 4.5 & 0 & 15 & 0\\
\rowcolor{gray!11}
\hspace{1em} \% of treated observations & 0.5 &  &  &  & \\
\addlinespace[0.3em]
\multicolumn{6}{l}{\textbf{Gender}}\\
\hspace{1em}Female & 23.1\% &  &  &  & 0\\
\hspace{1em}Male & 76.9\% &  &  &  & \\
\rowcolor{gray!11}
\addlinespace[0.3em]
\multicolumn{6}{l}{\textbf{Age}}\\
\rowcolor{gray!11}
\hspace{1em}Age at t & 46.6 & 8.3 & & & 96\\
\addlinespace[0.3em]
\multicolumn{6}{l}{\textbf{Institution Type}}\\
\hspace{1em}Cantonal university & 58.8\% &  &  &  & 1'439\\
\hspace{1em}ETH Domain & 23.9\% &  &  &  & \\
\hspace{1em}UAS/UTE/Other & 17.3\% &  &  &  & \\
\rowcolor{gray!11}
\addlinespace[0.3em]
\multicolumn{6}{l}{\textbf{Research Area}}\\
\rowcolor{gray!11}\hspace{1em}LS & 38.4\% &  &  &  & 0\\
\rowcolor{gray!11}
\hspace{1em}STEM & 31.6\% &  &  &  & \\
\rowcolor{gray!11}
\hspace{1em}SSH & 29.9\% &  &  &  & \\
\bottomrule
\end{tabular}
   \begin{threeparttable}
      \begin{tablenotes}
        \small  \item Notes: The data contains 82'249 researcher-year observations on 8'793 distinct researchers; \textit{av.} stands for average.
        \end{tablenotes} 
      \end{threeparttable}
\end{table}

\begin{figure} {\centering \includegraphics[width=1.05\linewidth]{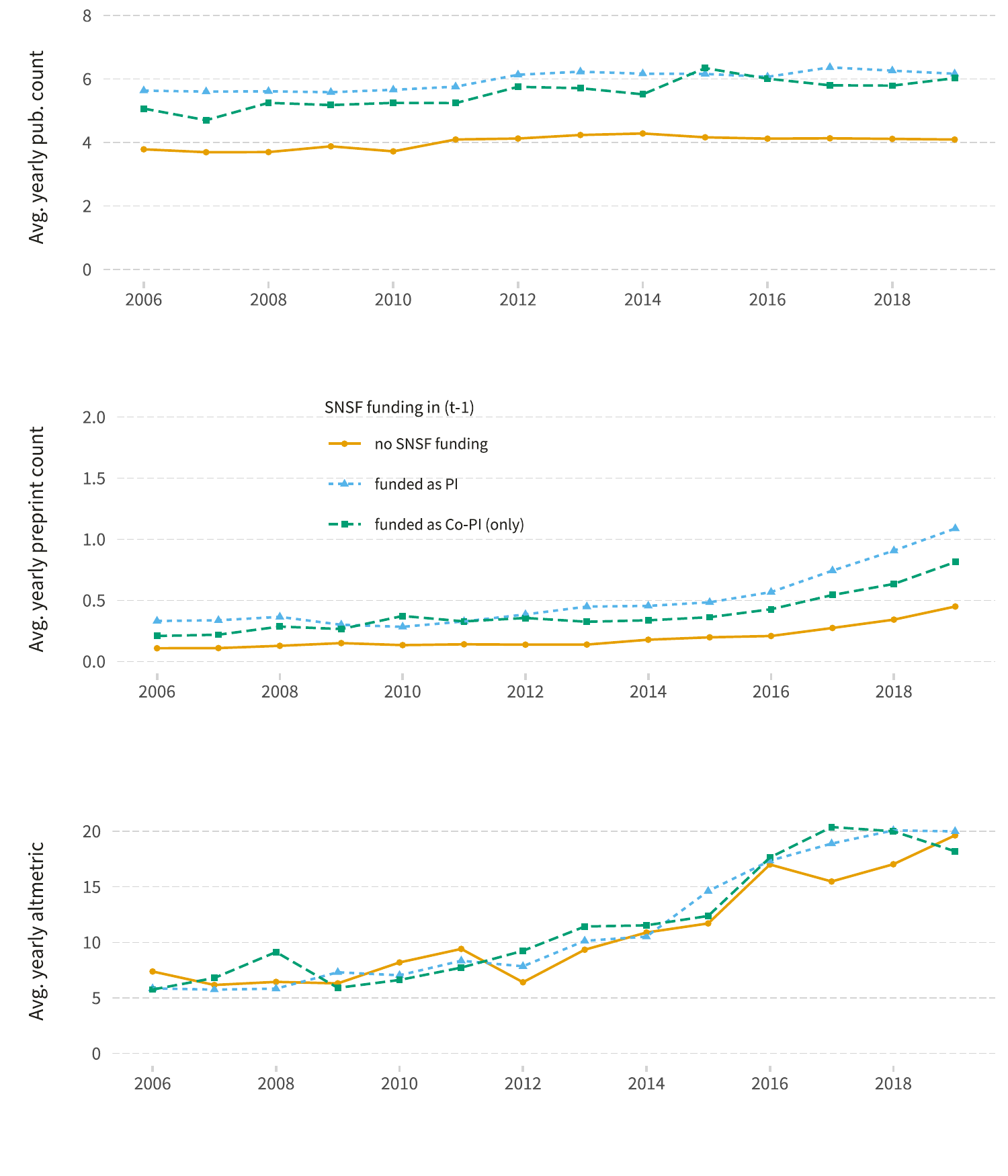}} \caption{Time trends of the publication and preprint counts as well as the altmetric score by SNSF funding status in the year before.}\label{fig:outcome_distribution}\end{figure}

Figure \ref{fig:outcome_distribution} represents the evolution of the yearly average number of articles, preprints and the altmetric score per researcher depending on the funding status of the year before (as co- and/or PI). The amount of articles published each year has been rather constant or only slightly increasing, while the preprint count increased substantially over the past years. Recent papers also have a higher altmetric scores than older publications, even though they had less time to raise attention.  
It is important to note, however, that since we do not account for any researcher characteristics here, the differences between funded and unfunded researchers cannot be interpreted as being the result of funding. Yet, increasing prevalence of preprints and altmetrics suggest that they should be taken into account in funding evaluations. 

\hypertarget{confounding-variables}{%
\subsubsection{Confounding variables}\label{confounding-variables}}

Table \ref{tab:output_baseline} further shows descriptive statistics for the gender of the researchers, their biological age, as well their field of research and the institution type. These variables capture drivers of researcher outputs and are therefore taken into account in all our analyses.
Almost 77\% of the researchers are male and about 60\% are employed at cantonal universities, 24\% at technical universities (ETH Domain) and about 17\% at University of Applied Sciences (UAS) and University of Teacher Education (UTE). 
The research field and institution type are defined as the area or the type the researcher applies most often to or from. The field of Life Science (LS) has the largest proposal share in the data with about 39\%. These variables serve as confounders together with the pre-sample information on the outcome variables since they may explain differences in output and therefore need to be accounted for. Note that 1'615 researchers in our data did not publish any peer-reviewed papers in the five year pre-sample period. Table \ref{tab:output_baseline_RA} in the Supplementary Material shows how the confounding variables vary between the research fields. 

The submitted project proposals are graded on a six-point scale: $1=D, 2=C, 3=BC, 4=B, 5=AB, 6=A$. 
We use the information on project evaluation to control for (or match on) average project quality following the approach by \citep{AroraGambardella_2005}. We construct the evaluation score as a rolling average over the last four years of all the grades a researcher `collected' in submitted proposals as PI and co-PI (if no grade was available over the last four years for a certain researcher, we use her all time average). We do so because future research is also impacted by the quality of past and co-occurring projects. 
The funding decision is, however, not exclusively based on those grades. It has to take the amount of funding available to the specific call into account. Therefore the ranking of an application among the other competing applications plays an important role and even highly rated projects may be rejected if the budget constraint is reached. Projects graded with an A/AB have good chances of being funded, while projects graded as D are never funded, see Figure \ref{fig:grade_distribution} representing the distribution of the grades among rejected and accepted projects.

Note that the researchers with missing age were deleted since this is an important control variable; the missing institution type were regrouped into unclassified. Additionally, for the analyses, the funding information will be used with a one (or more) year lag and at least one year of observation is lost per researcher. The final sample used for the analyses consists of 72'738 complete observations from 8'282 unique researchers.

\section{Results}\label{results}

\hypertarget{mixed-effects-model---longitudinal-regression-models}{%
\subsection{Mixed effects model - longitudinal regression models}\label{sec:mixed-models}}
Table \ref{tab:model_irr_count} summarises the results of both negative binomial mixed models for the count outcomes (yearly numbers of publications and preprints). The incidence rate ratios (IRR) inform us on the multiplicative change of the baseline count depending on funding status. The model for the publication count was fitted on the whole data set, while the model for the preprint count is fitted on data since 2010, because the number of preprints was rather small in general before. SNSF funding seems to have a significant positive effect on research productivity, regarding yearly publication counts (1.21 times higher for PI than without SNSF funding) as well as yearly preprint counts (1.30 times higher for the PI compared to researchers without SNSF funding).\footnote{Note that we also tested the robustness of this result to when focusing on PF as treatment and adding the researchers with a funded Sinergia project to the control group, but adjusting with a Sinergia dummy variable. The size of funding as PI and co-PI effects and their confidence intervals were comparable.} An `average' researcher without SNSF funding in $t-1$ publishes on average 4.64 articles in $t$. A similar researcher (with all confounding variables kept constant) with SNSF funding as PI in $t-1$ would publish 5.6 articles in $t$. 
PIs on an SNSF project publish more. The same is true for male researchers and younger researchers for preprints. Researchers from ETH Domain publish more than the ones from Cantonal Universities. Researchers publish more in recent years. Researchers in the LS publish more peer-reviewed articles compared to other research areas. Regarding preprints, we observe a different picture. Here STEM researchers publish more than researchers in LS. 

\begin{table}
\caption{\textbf{Incidence rate ratios} (IRR) of the multivariate \textbf{\textit{Negative Binomial}} models for the article and preprint counts}\label{tab:model_irr_count}
\small
\begin{tabular}[t]{lrllrll}
\toprule
\multicolumn{1}{c}{ } & \multicolumn{3}{c}{1. Articles} & \multicolumn{3}{c}{2. Preprints} \\
\multicolumn{1}{c}{ } & \multicolumn{3}{c}{(72'738 obs.)} & \multicolumn{3}{c}{(61'726 obs.)} \\
\cmidrule(l{3pt}r{3pt}){2-4} \cmidrule(l{3pt}r{3pt}){5-7}
  & IRR & (95\%-CI) & p-val. & IRR  & (95\%-CI)  & p-val. \\
\rowcolor{gray!6}
\midrule
Funded PI (t-1) & 1.21 & (1.19; 1.22) & $<$ 0.001 & 1.30 & (1.22; 1.39) & $<$ 0.001\\
\rowcolor{gray!6}
Funded Co-PI (t-1) & 1.11 & (1.09; 1.13) &  & 1.10 & (1.02; 1.19) & \\
Eval. score BC-B & 0.98 & (0.96; 1) & 0.026 & 0.79 & (0.74; 0.84) & $<$ 0.001\\
Eval. score C-D & 0.97 & (0.95; 0.99) &  & 0.55 & (0.49; 0.61) & \\
\rowcolor{gray!6}
Male (ref.: Female) & 1.46 & (1.37; 1.56) & $<$ 0.001 & 2.12 & (1.76; 2.56) & $<$ 0.001\\
Age (decades) at t & 1.01 & (0.99; 1.03) & 0.502 & 1.16 & (1.09; 1.24) & $<$ 0.001\\
\rowcolor{gray!6}
ETH Domain & 1.06 & (0.98; 1.14) & $<$ 0.001 & 1.44 & (1.21; 1.73) & $<$ 0.001\\
\rowcolor{gray!6}
UAS/UTE or Other & 0.56 & (0.52; 0.61) &  & 0.34 & (0.26; 0.45) & \\
\rowcolor{gray!6}
Unclassified & 1.13 & (1.04; 1.22) &  & 1.53 & (1.26; 1.86) & \\
Area STEM & 0.71 & (0.67; 0.77) & $<$ 0.001 & 3.10 & (2.63; 3.65) & $<$ 0.001\\
Area SSH & 0.15 & (0.14; 0.16) &  & 0.40 & (0.33; 0.49) & \\
\rowcolor{gray!6}
Year 2010-14 & 1.14 & (1.12; 1.16) & $<$ 0.001 &  &  & \\
\rowcolor{gray!6}
Year 2015-19 & 1.18 & (1.15; 1.21) &  &  &  & \\
\rowcolor{gray!6}
Year 2015-19 &  &  &  & 2.06 & (1.95; 2.17) & $<$ 0.001\\
\bottomrule
\end{tabular}
\begin{threeparttable}
      \begin{tablenotes}
        \small  \item Notes: Incidence Rate Ratios (IRR) provided together with their 95\% confidence intervals (CI). The p-values refer to the results of likelihood ratio tests for each variable to be present in the model. For the institution type University serves as reference category, for the fields it is Life Sciences. For the evaluation score the class AB-A serves as reference category and for year it is the period 06-09 for model 1 and 10-14 for model 2.
        \end{tablenotes} 
      \end{threeparttable}
\end{table}

Table \ref{tab:model_sum_ols_all} summarises the results of the four linear mixed models for the continuous outcomes: the average yearly number of citations per publication, the yearly average altmetric, the yearly average RCR and the yearly average FCR. Regarding the citation patterns, there is strong evidence that SNSF funding has a positive effect; especially PIs on SNSF projects have their articles cited more frequently (increase in average yearly citations of 0.33 per article for the PIs). Articles by LS researchers are cited most compared to researchers from other fields. This is also the case for researchers from ETH domain and older researchers. For altmetrics and citation ratios, we employ a logarithmic scale to account for the fact that their distributions are highly skewed; we can then interpret the coefficients as percentage change. Regarding altmetrics, research funded by the SNSF gets an attention score that is 5.1\% higher (by September 2020) compared to other researchers. Researchers in LS have by far the highest altmetrics followed by researchers in the SSH. There is no strong evidence for an effect of the funding on the average yearly RCR. This implies that in the short-run research outcomes of SNSF-funded researchers are as often cited as a mixed average of articles funded by the NIH or other important researcher funded world-wide, but also not significantly more than that. Younger researchers and researchers from the ETH domain have higher RCRs. The results also suggest a positive relation between SNSF funding and a researcher's FCR.

\begin{landscape}
\begin{table}
\caption{Coefficients and percentage changes from
        \textbf{\textit{normal linear mixed}} models for citation measures and altmetrics
        of a researcher.}
        \label{tab:model_sum_ols_all}
\centering
\resizebox{1.0\linewidth}{!}{
\begin{tabular}{lrlrrlrrlrrlr}
\toprule
\multicolumn{1}{c}{ } & \multicolumn{3}{c}{3. av. citations per publication} & \multicolumn{3}{c}{4. altmetric} & \multicolumn{3}{c}{5. RCR} & \multicolumn{3}{c}{6. FCR} \\
\multicolumn{1}{c}{ } & \multicolumn{3}{c}{(72'738 obs.)} & \multicolumn{3}{c}{(37'273 obs.)} & \multicolumn{3}{c}{(23'350 obs.)} & \multicolumn{3}{c}{(49'403 obs.)} \\
\cmidrule(l{3pt}r{3pt}){2-4} \cmidrule(l{3pt}r{3pt}){5-7} \cmidrule(l{3pt}r{3pt}){8-10} \cmidrule(l{3pt}r{3pt}){11-13}
  & Coef. estimate & 95\%-CI & p-val. & \% & 95\%-CI  & p-val.  & \%  & 95\%-CI   & p-val.   & \%   & 95\%-CI    & p-val.   \\
\rowcolor{gray!6}
\midrule
Funded PI (t-1) & 0.33 & (0.3; 0.4) & $<$ 0.001 & 5.1 & (1.7; 8.7) & 0.013 & 0.5 & (-1; 2.1) & 0.827 & 2.0 & (0.3; 3.8) & 0.031\\
\rowcolor{gray!6}
Funded Co-PI (t-1) & 0.23 & (0.2; 0.3) &  & 1.5 & (-2.5; 5.7) &  & 0.0 & (-1.9; 2) &  & 2.2 & (0.1; 4.3) & \\
Eval. score BC-B & -0.03 & (-0.1; 0) & $<$ 0.001 & -11.8 & (-15.3; -8.3) & $<$ 0.001 & -3.9 & (-6; -1.8) & $<$ 0.001 & -2.9 & (-4.8; -0.9) & $<$ 0.001\\
Eval. score C-D & -0.20 & (-0.3; -0.1) &  & -21.4 & (-25.3; -17.3) &  & -8.3 & (-10.6; -6) &  & -7.9 & (-10.2; -5.4) & \\
\rowcolor{gray!6}
Male & 0.18 & (0; 0.4) & 0.105 & 10.5 & (5; 16.3) & $<$ 0.001 & 1.5 & (-1.1; 4.2) & 0.256 & 8.3 & (5; 11.8) & $<$ 0.001\\
Age (decades) at t & 0.57 & (0.5; 0.7) & $<$ 0.001 & -2.2 & (-4.3; 0) & 0.053 & -3.1 & (-4.3; -2) & $<$ 0.001 & -11.2 & (-12.4; -10) & $<$ 0.001\\
\rowcolor{gray!6}
ETH Domain & 0.66 & (0.4; 0.9) & $<$ 0.001 & 14.3 & (8.1; 21) & $<$ 0.001 & 7.1 & (3.4; 11.1) & $<$ 0.001 & 8.8 & (5.1; 12.6) & $<$ 0.001\\
\rowcolor{gray!6}
UAS/UTE/Other & -0.96 & (-1.2; -0.7) &  & -3.1 & (-9.4; 3.8) &  & -5.4 & (-8.5; -2.3) &  & -9.8 & (-13.4; -6) & \\
\rowcolor{gray!6}
Unclassified & -0.01 & (-0.3; 0.3) &  & 14.2 & (7.5; 21.4) &  & 1.8 & (-1.4; 5.1) &  & 1.9 & (-1.9; 5.9) & \\
Area STEM & -1.68 & (-1.9; -1.4) & $<$ 0.001 & -36.9 & (-40; -33.6) & $<$ 0.001 &  &  &  & -23.7 & (-26.1; -21.3) & $<$ 0.001\\
Area SSH & -4.20 & (-4.4; -4) &  & -27.0 & (-31; -22.8) &  &  &  &  & -27.2 & (-29.5; -24.7) & \\
\rowcolor{gray!6}
Year 2010-14 & 0.76 & (0.7; 0.8) & $<$ 0.001 &  &  &  & 0.4 & (-1.1; 1.9) & 0.879 & -5.7 & (-7.4; -4.1) & $<$ 0.001\\
\rowcolor{gray!6}
Year 2015-19 & 1.26 & (1.2; 1.4) &  &  &  &  & 0.2 & (-1.5; 1.9) &  & -18.7 & (-20.3; -17.1) & \\
\rowcolor{gray!6}
Year 2015-19 &  &  &  & 51.1 & (47.4; 54.9) & $<$ 0.001 &  &  &  &  &  & \\
\bottomrule
\end{tabular}}
\begin{threeparttable}
   \begin{tablenotes}
      \small \item Notes: The coefficients of the models on the logarithmic scale of the outcomes are converted to percentage changes, while the coefficient for the average citation per publications reads as a unit change. The quantities are provided together with a 95\% confidence intervals (CI). For the model for the altmetric score, only data since 2010 is used, as this is a more recent metric. To model the RCR, only researchers from the LS were included so that we do not need to account for the research area.  The p-values refer to the results of likelihood ratio tests for each variable to be present in the model. For the institution type University serves as reference category, for the fields it is Life Sciences. For the evaluation score the class AB-A serves as reference category and for year it is the period 06-09 for model 3, 5, 6 and 10-14 for model 4.
    \end{tablenotes}
   \end{threeparttable}
\end{table}
\end{landscape}

\hypertarget{Non-Parametric}{%
\subsection{Non-Parametric Estimation}\label{non-parametric}}
While the previous estimation approaches modelled unobserved heterogeneity across individuals, the non-parametric matching approach addresses the selection into the treatment explicitly. It accounts for selection on observable factors which may - if not accounted for - lead to wrongly attributing the funding effect to the selectivity of the grant-awarding process. We model a researcher's funding success 
as a function of researcher characteristics. In particular, this includes their previous research track record (publication experience and citations) and the average of all evaluation scores for submitted proposals (PI or co-PI) received by the researcher. In addition, we include age,  gender, research field and institution type. We obtain the propensity score to be used in the matching process as described in Section \ref{subsec:empirical-model}. 

The results from the probit estimation on the funding outcome (success vs. rejection) are presented in Table \ref{tab:probit}. The table first shows the model for the full sample which provides the propensity score for the estimation of treatment effects on articles and citations to these articles, and on preprints. The second model shows the model for the sub-sample of researchers in the LS used for estimating treatment effects on the RCR. The third model shows the estimation for the full sample, but accounting for pre-sample FCR, and provides the propensity score for the estimation of the treatment effect on the FCR. The fourth model controls for pre-sample altmetrics values and serves the estimation of the treatment effect on future altmetrics scores. 
Consistent across all specification, the results show that the evaluation score is a key predictor of grant success. The higher the score, the more likely is it that a proposal gets approved. The grant likelihood for male researches is higher than for females as well as for older researchers. The latter result can have various reasons, which are outside the scope of this paper and are being discussed elsewhere\footnote{\cite{Severin2020}, for example, discuss gender biases on the reviewer scores leading to lower grant likelihood for female researchers.}. As expected, past research performance is another strong predictor of grant success where peer-reviewed articles matter more than preprints. In addition to quantity, past research quality (as measured by citations) increases the probability of a proposal being granted. Interesting in more recent years (as shown in model 4), quality rather than quantity appears to predict grant success as it is the average number of citations to pre-period publication rather than their number that explains funding success. 

The comparison of the distribution of the propensity score and the evaluation score before and after matching shows that the nearest neighbor matching procedure was successful in balancing the sample in terms of the grant likelihood and - importantly - also the average scores (see Figure \ref{fig:matchingsuccess}). This ensures that we are comparing researchers with funding to researchers without funding that have similarly good ideas (the scores are the same, on average) and are also otherwise comparable in their characteristics predicting a positive application outcome. The balancing of the propensity scores and the evaluation scores in both groups (grant winners and unsuccessful applicants) after each matching are shown in Tables \ref{tab:matchedpub} and \ref{tab:treatment2}. Note that we draw matches for each grant-winner from the control group with replacement and that hence some observations from researchers in the control group are used several times as `twins'. Table \ref{tab:drawing} shows that across the different matched samples less than 10\% of control researcher-year observations are used only once and about 60\% up to 25 times. About 10\% of control group researchers are used very frequently, i.e. more than 160 times. 

Tables \ref{tab:matchedpub} and \ref{tab:treatment2} show the estimated treatment effects after matching, i.e. the test for the magnitude and significance of mean differences across groups. Note that the number of matched pairs differs depending on the sample used and that log values of output variables were used to account for the impact of skewness of the raw variable distribution in the mean comparison test. 
The magnitude of the estimated effects is comparable to the ones of the parametric estimation models. Researchers with a successful grant publish on average 1.2 articles (exp[0.188]) and about one additional preprint  (exp[0.053]) more in the following year, their articles receive 1.7 citations  (exp(0.532)) more than articles from the control group. In terms of altmetrics we also see a significant difference in means which is 1.15  (exp[0.138]) points higher in the group of grant receivers. Also, in terms of the FCR and the RCR, there are significant effects on the treatment group. The probability to be among the `highly cited researchers' (as measured by an FCR $>$3) is 5.5 ($\alpha_{TT}=0.055$) percentage points higher in the group of funded researchers. This means publications in $t+1$ are cited at least three times as much as the average in the field.

\begin{landscape}
\begin{table}
  \centering \small
  \caption{Estimation of \textbf{\textit{probit models}} on the funding propensity as PI or Co-PI }
  \resizebox{1.0\linewidth}{!}{
\begin{tabular}{ld{2.0}d{2.0}d{2.0}d{2.0}d{2.0}d{2.0}d{2.0}d{2.0}d{2.0}d{2.0}d{2.0}d{2.0}d{2.0}}
    \toprule
          & \multicolumn{3}{c}{1. articles} & \multicolumn{3}{c}{2. RCR } & \multicolumn{3}{c}{3. FCR} & \multicolumn{3}{c}{4. altmetrics} \\
          & \multicolumn{3}{c}{(All Fields)} & \multicolumn{3}{c}{(Life Sciences)} & \multicolumn{3}{c}{(All Fields)} & \multicolumn{3}{c}{(All since 2010)} \\
          & \multicolumn{1}{r}{Coef.} & \multicolumn{1}{r}{SE} & \multicolumn{1}{r}{$P>|z|$} & \multicolumn{1}{r}{Coef.} & \multicolumn{1}{r}{SE} & \multicolumn{1}{r}{$P>|z|$} & \multicolumn{1}{r}{Coef.} & \multicolumn{1}{r}{SE} & \multicolumn{1}{r}{$P>|z|$} & \multicolumn{1}{r}{Coef.} & \multicolumn{1}{r}{SE} & \multicolumn{1}{r}{$P>|z|$} \\
    \midrule
    \rowcolor{gray!11}
    ln(pre-sample articles) & \multicolumn{1}{r}{0.053} & \multicolumn{1}{r}{0.012} & \multicolumn{1}{r}{$<$0.001} & \multicolumn{1}{r}{0.050} & \multicolumn{1}{r}{0.020} & \multicolumn{1}{r}{0.012} & \multicolumn{1}{r}{0.029} & \multicolumn{1}{r}{0.014} & \multicolumn{1}{r}{0.038} & \multicolumn{1}{r}{-0.013} & \multicolumn{1}{r}{0.018} & \multicolumn{1}{r}{0.484} \\
    \rowcolor{gray!11}
    
    ln(pre-sample preprints) & \multicolumn{1}{r}{0.006} & \multicolumn{1}{r}{0.028} & \multicolumn{1}{r}{0.840} & \multicolumn{1}{r}{0.164} & \multicolumn{1}{r}{0.159} & \multicolumn{1}{r}{0.300} & \multicolumn{1}{r}{-0.003} & \multicolumn{1}{r}{0.030} & \multicolumn{1}{r}{0.930} & \multicolumn{1}{r}{0.060} & \multicolumn{1}{r}{0.037} & \multicolumn{1}{r}{0.100} \\
    
        ln(pre-sample av. citations) & \multicolumn{1}{r}{0.034} & \multicolumn{1}{r}{0.008} & \multicolumn{1}{r}{$<$0.001} &       &       &       &       &       &       & \multicolumn{1}{r}{0.067} & \multicolumn{1}{r}{0.013} & \multicolumn{1}{r}{$<$0.001} \\

    ln(pre-sample RCR) &       &       &       & \multicolumn{1}{r}{0.051} & \multicolumn{1}{r}{0.021} & \multicolumn{1}{r}{0.013} &       &       &       &       &       &  \\
    ln(pre-sample FCR) &       &       &       &       &            &   & \multicolumn{1}{r}{0.040} & \multicolumn{1}{r}{0.009} & \multicolumn{1}{r}{$<$0.001} &       &       &  \\
    \rowcolor{gray!11}
    ln(pre-sample altmetrics) &       &       &       &       &       &       & \multicolumn{1}{r}{} & \multicolumn{1}{r}{} & \multicolumn{1}{r}{} & \multicolumn{1}{r}{-0.089} & \multicolumn{1}{r}{0.012} & \multicolumn{1}{r}{$<$0.001} \\

    Eval.score & \multicolumn{1}{r}{0.815} & \multicolumn{1}{r}{0.007} & \multicolumn{1}{r}{$<$0.001} & \multicolumn{1}{r}{0.760} & \multicolumn{1}{r}{0.010} & \multicolumn{1}{r}{$<$0.001} & \multicolumn{1}{r}{0.798} & \multicolumn{1}{r}{0.008} & \multicolumn{1}{r}{$<$0.001} & \multicolumn{1}{r}{0.800} & \multicolumn{1}{r}{0.010} & \multicolumn{1}{r}{$<$0.001} \\
    \rowcolor{gray!11}
    
    Male researcher & \multicolumn{1}{r}{0.120} & \multicolumn{1}{r}{0.016} & \multicolumn{1}{r}{$<$0.001} & \multicolumn{1}{r}{0.135} & \multicolumn{1}{r}{0.027} & \multicolumn{1}{r}{$<$0.001} & \multicolumn{1}{r}{0.111} & \multicolumn{1}{r}{0.019} & \multicolumn{1}{r}{$<$0.001} & \multicolumn{1}{r}{0.102} & \multicolumn{1}{r}{0.026} & \multicolumn{1}{r}{$<$0.001} \\
    \rowcolor{gray!11}
    
    ln(age) & \multicolumn{1}{r}{1.703} & \multicolumn{1}{r}{0.038} & \multicolumn{1}{r}{$<$0.001} & \multicolumn{1}{r}{2.041} & \multicolumn{1}{r}{0.068} & \multicolumn{1}{r}{$<$0.001} & \multicolumn{1}{r}{1.891} & \multicolumn{1}{r}{0.045} & \multicolumn{1}{r}{$<$0.001} & \multicolumn{1}{r}{1.972} & \multicolumn{1}{r}{0.062} & \multicolumn{1}{r}{$<$0.001} \\
    
    ETH Domain & \multicolumn{1}{r}{0.066} & \multicolumn{1}{r}{0.020} & \multicolumn{1}{r}{$<$0.001} & \multicolumn{1}{r}{0.156} & \multicolumn{1}{r}{0.041} & \multicolumn{1}{r}{$<$0.001} & \multicolumn{1}{r}{0.047} & \multicolumn{1}{r}{0.023} & \multicolumn{1}{r}{0.038} & \multicolumn{1}{r}{0.059} & \multicolumn{1}{r}{0.030} & \multicolumn{1}{r}{0.050} \\
    
    UAS/UTE/Other & \multicolumn{1}{r}{-0.271} & \multicolumn{1}{r}{0.018} & \multicolumn{1}{r}{$<$0.001} & \multicolumn{1}{r}{-0.243} & \multicolumn{1}{r}{0.033} & \multicolumn{1}{r}{$<$0.001} & \multicolumn{1}{r}{-0.294} & \multicolumn{1}{r}{0.024} & \multicolumn{1}{r}{$<$0.001} & \multicolumn{1}{r}{-0.268} & \multicolumn{1}{r}{0.032} & \multicolumn{1}{r}{$<$0.001} \\
    
    Unclassified  & \multicolumn{1}{r}{-0.521} & \multicolumn{1}{r}{0.021} & \multicolumn{1}{r}{$<$0.001} & \multicolumn{1}{r}{-0.406} & \multicolumn{1}{r}{0.034} & \multicolumn{1}{r}{$<$0.001} & \multicolumn{1}{r}{-0.580} & \multicolumn{1}{r}{0.023} & \multicolumn{1}{r}{$<$0.001} & \multicolumn{1}{r}{-0.635} & \multicolumn{1}{r}{0.029} & \multicolumn{1}{r}{$<$0.001} \\
    
    \rowcolor{gray!11}
    STEM & \multicolumn{1}{r}{-0.138} & \multicolumn{1}{r}{0.019} & \multicolumn{1}{r}{$<$0.001} &       &       &       & \multicolumn{1}{r}{-0.098} & \multicolumn{1}{r}{0.021} & \multicolumn{1}{r}{$<$0.001} & \multicolumn{1}{r}{-0.118} & \multicolumn{1}{r}{0.027} & \multicolumn{1}{r}{$<$0.001} \\
    \rowcolor{gray!11}
    
    SSH   & \multicolumn{1}{r}{0.019} & \multicolumn{1}{r}{0.019} & \multicolumn{1}{r}{0.307} &       &       &       & \multicolumn{1}{r}{0.053} & \multicolumn{1}{r}{0.023} & \multicolumn{1}{r}{0.023} & \multicolumn{1}{r}{0.114} & \multicolumn{1}{r}{0.033} & \multicolumn{1}{r}{$<$0.001} \\
    
\cmidrule{1-13}    JS Inst. types  & \multicolumn{2}{c}{889.44} & \multicolumn{1}{r}{$<$0.001} & \multicolumn{2}{c}{212.92} & \multicolumn{1}{r}{$<$0.001} &      \multicolumn{2}{c}{802.72}      & \multicolumn{1}{r}{$<$0.001}      & \multicolumn{2}{c}{890.49} & \multicolumn{1}{r}{$<$0.001} \\

\rowcolor{gray!11}
    JS fields & \multicolumn{2}{c}{71.42} & \multicolumn{1}{r}{$<$0.001} & \multicolumn{3}{c}{}  &   \multicolumn{2}{c}{39.26}    &  \multicolumn{1}{r}{$<$0.001}        & \multicolumn{2}{c}{ 72.62} & \multicolumn{1}{r}{$<$0.001} \\
    
    JS years  & \multicolumn{2}{c}{1'469.08} & \multicolumn{1}{r}{$<$0.001} & \multicolumn{2}{c}{797.32} & \multicolumn{1}{r}{$<$0.001} & \multicolumn{2}{c}{1'195.22}        &  \multicolumn{1}{r}{$<$0.001}     & \multicolumn{2}{c}{1'551.61} & \multicolumn{1}{r}{$<$0.001} \\
    \rowcolor{gray!11}
    \# observations & \multicolumn{3}{c}{63'680} & \multicolumn{3}{c}{22'999} & \multicolumn{3}{c}{48'729} & \multicolumn{3}{c}{30'360} \\
    
    Log pseudolikelihood & \multicolumn{3}{c}{-24'863.80} & \multicolumn{3}{c}{-9'369.45} & \multicolumn{3}{c}{-18'226.64} & \multicolumn{3}{c}{-10'896.19} \\
    \rowcolor{gray!11}
    Pseudo $R^2$ & \multicolumn{3}{c}{0.361} & \multicolumn{3}{c}{0.341} & \multicolumn{3}{c}{0.363} & \multicolumn{3}{c}{0.367} \\
    \midrule
     \end{tabular}}
     \begin{threeparttable}
      \begin{tablenotes}
      \small
      \item Notes: The models contain a constant and year dummies (not reported). JS stands for $chi^2$-test of joint significance of the respective factor. For the institution type University serves as reference category, for the fields it is Life Sciences. The specification for preprints is identical to model 1, but for observations since the year 2010 only (n = 52'410 Pseudo $R^2$ = 38.8).  
    \end{tablenotes}
   \end{threeparttable}
  \label{tab:probit}%
\end{table}%
\end{landscape}

\hypertarget{persistency}{%
\subsection{Persistency of treatment effects}\label{persistency}}
In addition to the effect in the year after funding $(t+1)$, we are interested in the persistency of the effect in the following years up to $(t+3)$. 
It is likely that any output effects occur with a considerable time-lag after funding received. The start-up of the research project including the training of new researchers and the set-up of equipment may take some time before the actual research starts.
In principle, we could of course expect the effect to last also longer than three to four years. However, after four years, the treatment effect of one project grant may become confounded by one (or several) follow-up grants. 
Tables \ref{tab:matchedpub} and \ref{tab:treatment2} show the results for the different outcome variables also for different time horizons. 

The results suggest that the funding has a persistent output effect amounting to about one additional article in each of the three years following the year of funding.  The effect on preprints is already significant in the first year, but also turns out to sustain in later years suggesting that research results from the project are probably circulated via this channel. In contrast to these results, we find for altmetrics that they are significantly higher early on, but not in the medium-run.
When looking at citation-based measures as indicators for impact and relevance, we see that the number of citations stays significantly higher in the medium-run, but effect size declines somewhat indicating that researchers publish the most important results earlier after funding. This is also reflected in the results for the average number of citations and the probability to be highly cited.  For the FCR, the effect is less persistent as the difference between groups fades after the first year. For the RCR the differences in means is strongest in the first year after the grant and only significant at the 10\% level in {t+3}.

\begin{landscape}
\begin{table}
  \centering 
  \caption{Treatment effects (after matching): articles and citations}
    \resizebox{\linewidth}{!}{
    \begin{tabular}{lrrrrrr}
      \toprule
                   & ln(articles+1)$_{t+1}$ & ln(articles+1)$_{t+2}$ & ln(articles+1)$_{t+3}$ & ln(citations+1)$_{t+1}$ & ln(citations+1)$_{t+2}$ & ln(citations+1)$_{t+3}$ \\
    \midrule
      \rowcolor{gray!11}
    Treated        & \multicolumn{1}{r}{1.409} & \multicolumn{1}{r}{1.421} & \multicolumn{1}{r}{1.431} & \multicolumn{1}{r}{3.756} & \multicolumn{1}{r}{3.909} & \multicolumn{1}{r}{4.038} \\
      \rowcolor{gray!11}
    SE             & \multicolumn{1}{r}{0.005} & \multicolumn{1}{r}{0.005} & \multicolumn{1}{r}{0.005} & \multicolumn{1}{r}{0.010} & \multicolumn{1}{r}{0.011} & \multicolumn{1}{r}{0.011} \\
    Control        & \multicolumn{1}{r}{1.221} & \multicolumn{1}{r}{1.227} & \multicolumn{1}{r}{1.274} & \multicolumn{1}{r}{3.223} & \multicolumn{1}{r}{3.482} & \multicolumn{1}{r}{3.670} \\
    SE             & \multicolumn{1}{r}{0.005} & \multicolumn{1}{r}{0.005} & \multicolumn{1}{r}{0.005} & \multicolumn{1}{r}{0.009} & \multicolumn{1}{r}{0.010} & \multicolumn{1}{r}{0.010} \\
      \rowcolor{gray!11}
    $\alpha_{TT}$  & \multicolumn{1}{r}{0.188} & \multicolumn{1}{r}{0.194} & \multicolumn{1}{r}{0.157} & \multicolumn{1}{r}{0.533} & \multicolumn{1}{r}{0.427} & \multicolumn{1}{r}{0.368} \\  
    \rowcolor{gray!11}
    Diff. interval & \multicolumn{1}{r}{0.175-0.201} & \multicolumn{1}{r}{0.180-0.207} & \multicolumn{1}{r}{0.142-0.171} & \multicolumn{1}{r}{0.505-0.559} &   \multicolumn{1}{r}{0.398-0.455} & \multicolumn{1}{r}{0.337-0.398} \\
      \rowcolor{gray!11}
    Pr($|T| > |t|$)& \multicolumn{1}{r}{$<$0.001} & \multicolumn{1}{r}{$<$0.001} & \multicolumn{1}{r}{$<$0.001} & \multicolumn{1}{r}{$<$0.001} & \multicolumn{1}{r}{$<$0.001} & \multicolumn{1}{r}{$<$0.001} \\
    \midrule
    PS Treated     & \multicolumn{1}{r}{0.819} &       &       & \multicolumn{1}{r}{0.819} &       &  \\
    PS Control     & \multicolumn{1}{r}{0.813} &       &       & \multicolumn{1}{r}{0.813} &       &  \\
    Pr($|T| > |t|$) & \multicolumn{1}{r}{0.313} &       &       & \multicolumn{1}{r}{0.313} &       &  \\
      \rowcolor{gray!11}
    ES Treated      & \multicolumn{1}{r}{3.929} &       &       & \multicolumn{1}{r}{3.929} &       &  \\
      \rowcolor{gray!11}
    ES Control      & \multicolumn{1}{r}{3.904} &       &       & \multicolumn{1}{r}{3.904} &       &  \\
      \rowcolor{gray!11}
    Pr($|T| > |t|$) & \multicolumn{1}{r}{0.405} &       &       & \multicolumn{1}{r}{0.405} &       &  \\
    \# matched  & \multicolumn{1}{r}{43'936} & \multicolumn{1}{r}{39'125} & \multicolumn{1}{r}{34'383} & \multicolumn{1}{r}{43'936} & \multicolumn{1}{r}{39'125} & \multicolumn{1}{r}{34'383} \\

    \midrule
          & ln(av.citations+1)$_{t+1}$ & ln(av.citations+1)$_{t+2}$ & ln(av.citations+1)$_{t+3}$ & highly cited$_{t+1}$ & highly cited$_{t+1}$ & highly cited$_{t+1}$ \\
    \midrule
      \rowcolor{gray!11}
    Treated    & \multicolumn{1}{r}{1.488} & \multicolumn{1}{r}{1.521} & \multicolumn{1}{r}{1.546} & \multicolumn{1}{r}{0.673} & \multicolumn{1}{r}{0.666} & \multicolumn{1}{r}{0.658} \\
      \rowcolor{gray!11}
    SE         & \multicolumn{1}{r}{0.004} & \multicolumn{1}{r}{0.004} & \multicolumn{1}{r}{0.004} & \multicolumn{1}{r}{0.003} & \multicolumn{1}{r}{0.003} & \multicolumn{1}{r}{0.003} \\
    Control    & \multicolumn{1}{r}{1.402} & \multicolumn{1}{r}{1.480} & \multicolumn{1}{r}{1.518} & \multicolumn{1}{r}{0.618} & \multicolumn{1}{r}{0.618} & \multicolumn{1}{r}{0.656} \\
    SE         & \multicolumn{1}{r}{0.004} & \multicolumn{1}{r}{0.004} & \multicolumn{1}{r}{0.004} & \multicolumn{1}{r}{0.003} & \multicolumn{1}{r}{0.003} & \multicolumn{1}{r}{0.003} \\
      \rowcolor{gray!11}
    $\alpha_{TT}$ & \multicolumn{1}{r}{0.086} & \multicolumn{1}{r}{0.041} & \multicolumn{1}{r}{0.028} & \multicolumn{1}{r}{0.055} & \multicolumn{1}{r}{0.048} & \multicolumn{1}{r}{0.002} \\
      \rowcolor{gray!11}
    Diff. interval & \multicolumn{1}{r}{0.075-0.097} & \multicolumn{1}{r}{0.030-0.052} & \multicolumn{1}{r}{0.016-0.039} & \multicolumn{1}{r}{0.047-0.062} & \multicolumn{1}{r}{0.039-0.056} & \multicolumn{1}{r}{0.007-0.011} \\
      \rowcolor{gray!11}
    Pr($|T| > |t|$) & \multicolumn{1}{r}{$<0.001$} & \multicolumn{1}{r}{0.109} & \multicolumn{1}{r}{0.327} & \multicolumn{1}{r}{0.007} & \multicolumn{1}{r}{0.042} & \multicolumn{1}{r}{0.941} \\
    \midrule
    PS Treated & \multicolumn{1}{r}{0.819} &       &       & \multicolumn{1}{r}{0.833} &       &  \\
    PS Control & \multicolumn{1}{r}{0.813} &       &       & \multicolumn{1}{r}{0.826} &       &  \\
    Pr($|T| > |t|$) & \multicolumn{1}{r}{0.313} &       &       & \multicolumn{1}{r}{0.272} &       &  \\
      \rowcolor{gray!11}
    ES Treated & \multicolumn{1}{r}{3.929} &       &       & \multicolumn{1}{r}{3.955} &       &  \\
      \rowcolor{gray!11}
    ES Control & \multicolumn{1}{r}{3.904} &       &       & \multicolumn{1}{r}{3.930} &       &  \\
      \rowcolor{gray!11}
    Pr($|T| > |t|$) & \multicolumn{1}{r}{0.405} &       &       & \multicolumn{1}{r}{0.476} &       &  \\
    \# matched  & \multicolumn{1}{r}{43'936} & \multicolumn{1}{r}{39'125} & \multicolumn{1}{r}{34'383} & \multicolumn{1}{r}{28'936} & \multicolumn{1}{r}{25'436} & \multicolumn{1}{r}{22'059} \\
    \midrule
 \end{tabular}}     
   \begin{threeparttable}
   \begin{tablenotes}
      \small \item Notes: The table shows sample means and results from t-tests on mean differences after matching. PS stands for Propensity Score and ES stands for Evaluation Score (which is here used as a numeric value). The number of matched pairs declines with each lead due to censoring at the end of the observation period.
    \end{tablenotes}
   \end{threeparttable}
   \label{tab:matchedpub} 
\end{table}
 \end{landscape}

\begin{landscape}
\begin{table}
  \centering 
  \caption{Treatment effects (after matching): preprints, altmetrics and citation rates}
   \resizebox{\linewidth}{!}{
    \begin{tabular}{lrrrrrr}
    \toprule
          & \multicolumn{1}{r}{ln(preprints+1)$_{t+1}$} & \multicolumn{1}{r}{ln(preprints+1)$_{t+2}$} & \multicolumn{1}{r}{ln(preprints+1)$_{t+3}$} & \multicolumn{1}{r}{ln(altmetrics+1)$_{t+1}$} & \multicolumn{1}{r}{ln(altmetrics+1)$_{t+2}$} & \multicolumn{1}{r}{ln(altmetrics+1)$_{t+3}$} \\
    \midrule
     \rowcolor{gray!11}
    Treated  & 0.203 & 0.209 & 0.215 & 2.010 & 2.079 & 2.103 \\
     \rowcolor{gray!11}
    SE       & 0.003 & 0.003 & 0.003 & 0.007 & 0.008 & 0.009 \\
    Control  & 0.150 & 0.174  & 0.174 & 1.872 & 2.028 & 1.997 \\
    SE       & 0.002 & 0.003 & 0.003 & 0.007 & 0.008 & 0.009 \\
     \rowcolor{gray!11}
    $\alpha_{TT}$ & 0.053 & 0.034 & 0.040 & 0.138 & 0.051 & 0.106 \\
     \rowcolor{gray!11}
    Diff. interval & 0.046-0.059 & 0.027-0.042 & 0.032-0.049 & 0.120-0.157 & 0.029-0.073 &  0.082-0.130 \\
     \rowcolor{gray!11}
    Pr($|T| > |t|$) & $<$0.001 & 0.044 & 0.041 & $<$0.001 & 0.368 & 0.108 \\
    \midrule
    PS Treated & 0.819 &       &       & 0.845 &       &  \\
    PS Control & 0.814 &       &       & 0.836 &       &  \\
    Pr($|T| > |t|$) & 0.433  &       &       & 0.187 &       &  \\
     \rowcolor{gray!11}
    ES Treated & 3.920 &       &       & 3.951 &       &  \\
     \rowcolor{gray!11}
    ES Control & 3.899 &       &       & 3.906 &       &  \\
     \rowcolor{gray!11}
    Pr($|T| > |t|$) & 0.507 &       &       & 0.274 &       &  \\
    \# matched  & 35'330 & 30'519 & 25'776 & 22'362 & 17'615 & 15'129 \\
 
    \midrule
          & \multicolumn{1}{r}{ln(FCR+1)$_{t+1}$} & \multicolumn{1}{r}{ln(FCR+1)$_{t+2}$} & \multicolumn{1}{r}{ln(FCR+1)$_{t+3}$} & \multicolumn{1}{r}{ln(RCR+1)$_{t+1}$} & \multicolumn{1}{r}{ln(RCR+1)$_{t+2}$} & \multicolumn{1}{r}{ln(RCR+1)$_{t+3}$} \\
    \midrule
     \rowcolor{gray!11}
    Treated  & 1.700 & 1.684 & 1.669 & 0.885 & 0.883 & 0.880 \\
     \rowcolor{gray!11}
    SE    & 0.004 & 0.005 & 0.005 & 0.004 & 0.004 & 0.004 \\
    Control  & 1.635 & 1.666 & 1.676 & 0.827 & 0.837 & 0.832 \\
    SE    & 0.005 & 0.005 & 0.005 & 0.004 & 0.004 & 0.004 \\
     \rowcolor{gray!11}
    $\alpha_{TT}$ & 0.065 & 0.018 & -0.007 & 0.058 & 0.046 & 0.048 \\
     \rowcolor{gray!11}
    Diff. interval & 0.052-0.072 & -0.007-0.020 & 0.012-0.014 & 0.047-0.069 & 0.035-0.057 & 0.036-0.060 \\
     \rowcolor{gray!11}
    Pr($|T| > |t|$) & 0.042 & 0.614 & 0.869 & 0.009 & 0.060 & 0.079 \\
    \midrule
    PS Treated & 0.833 &       &       & 0.802 &       &  \\
    PS Control & 0.826 &       &       & 0.793 &       &  \\
    Pr($|T| > |t|$) & 0.272  &       &       & 0.306 &       &  \\
     \rowcolor{gray!11}
    ES Treated & 3.955 &       &       & 3.659 &       &  \\
     \rowcolor{gray!11}
    ES Control & 3.930 &       &       & 3.630 &       &  \\
     \rowcolor{gray!11}
    Pr($|T| > |t|$) & 0.476 &       &       & 0.508 &       &  \\
    \# matched  & 28'936 & 25'436 & 22'059 & 13'244 & 11'752 & 10'312 \\
    \midrule
  \end{tabular}}%
\label{tab:treatment2}    
\begin{threeparttable}  
\begin{tablenotes}
      \small \item Notes: The table shows sample means and results from t-tests on mean differences after matching. PS stands for Propensity Score and ES stands for Evaluation Score (which is here used as a numeric value). The number of matched pairs declines with each lead due to censoring at the end of the observation period. The treatment effect on RCR is estimated only for the Life Sciences and for altmetrics only since 2010.
    \end{tablenotes}
   \end{threeparttable}
\end{table}
\end{landscape}

\subsection{Impact heterogeneity over the academic life-cycle and research fields}
For most outcomes, we find a significant and persistent difference between funded and unfunded researchers, while controlling for other drivers of research outcomes. As shown in earlier studies \citep{AroraGambardella_2005,JacobLefgren_2011}, a grant's impact may depend on the career stage of a researcher. As a proxy for career stage, we use the biological age of the researchers. Additionally, there might be heterogeneity in the funding effect depending on the research fields. We perform interaction tests between (i) the age and the funding and (ii) between the research field and the funding. More specifically, we employ a categorical variable for age and allow for an interaction term with the funding variable in the mixed models presented in Section \ref{sec:mixed-models}. The same procedure is repeated with research field. The interaction tests suggests indeed that there is evidence for a difference in the effect of funding on the article and preprint count depending on the age group (with p-value $<$ 0.001, for both outcomes) and the research field (with p-value of $<$ 0.001 for articles and p-value of 0.0045 for preprints). When we test for those same interaction effects in the continuous outcome models, the results suggest that there is a difference in the funding effect on the average number of citations per article depending on the age group (p-value $<$ 0.001) and the research field (p-value = 0.0242). For altmetrics and the citation ratios, we see no evidence for major differences across age groups (p-value of 0.328 for the altmetric, 0.802 for the RCR and 0.873 for the FCR) nor research fields (p-value of 0.2296 for altmetric and p-value of 0.5124 for FCR\footnote{Note that we did not test the interaction for the RCR outcome, as this analysis was done only for the LS field.}).

\begin{figure} \centering
  \includegraphics[width=1\linewidth]{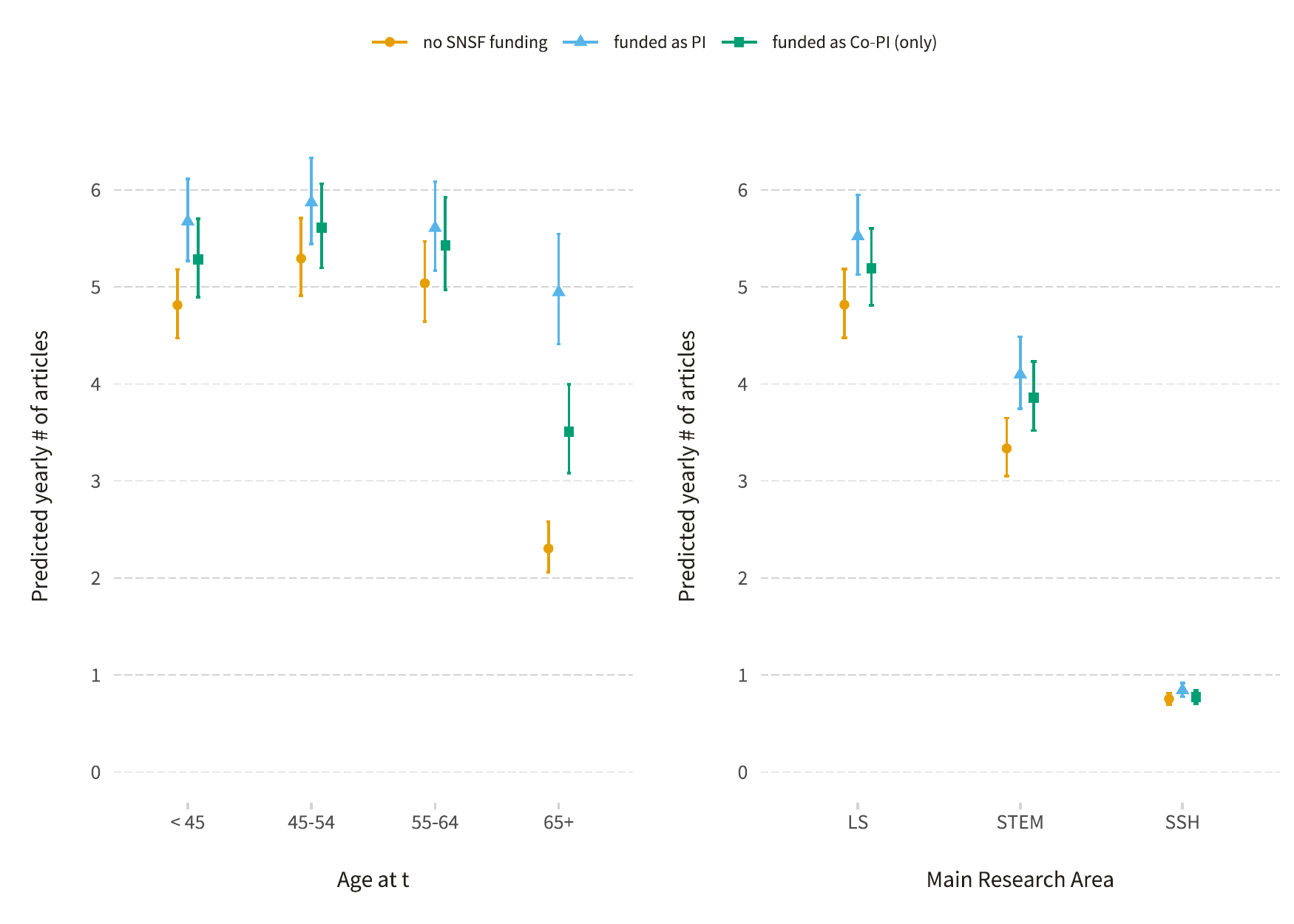} \label{fig:interactions_articles}
  \caption{Predicted yearly number of articles depending on whether or not the researchers were treated (as PI or Co-PI) or not (no funding), given the researcher's age group (left) and the research field (right). \\
  {\scriptsize To predict the article count the baseline confounding variables were fixed to Year 2015-19, Male, Evaluation Score Score AB-A, University, LS in the age interaction model and age lower to 45 for the field interaction model.
  We see a significant positive percentage change of 78\% for the youngest age group among PIs ($<$ 45) and 115\% for the most senior researchers ($>$ 65) compared to no SNSF funding. Additionally, the effect of funding is largest for STEM researchers (23\% more articles as PI compared to unfunded researchers. The effect in LS and SSH is less prominent, +15\% and +12\% respectively.}}\end{figure}

\begin{figure} \centering
  \includegraphics[width=1\linewidth]{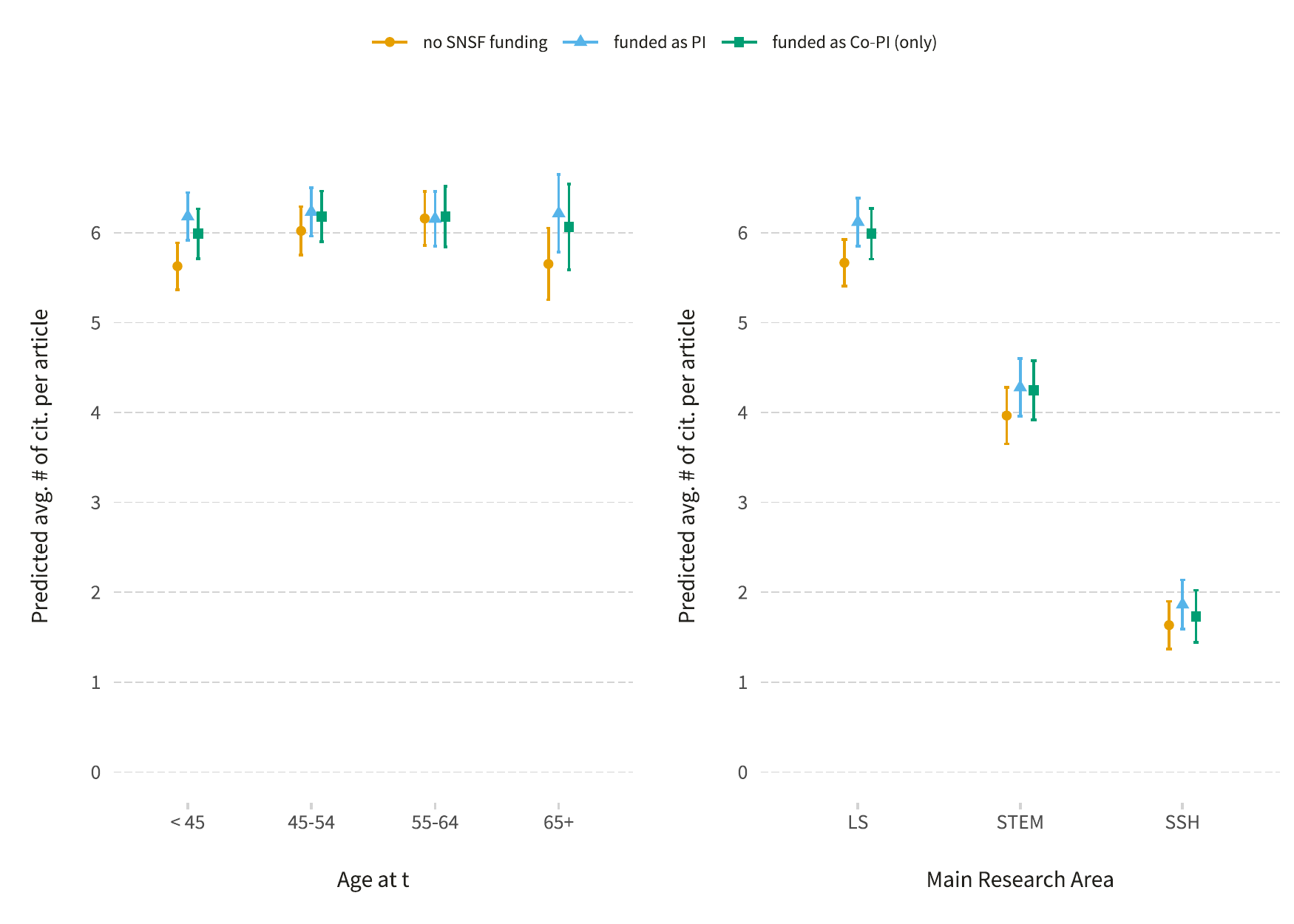} \label{fig:interactions_citations}
  \caption{Predicted yearly average number of citations per article depending on whether or not the researchers were treated or not, given the researcher's age group (left) and the research field (right).\\
   {\scriptsize For the predictions the baseline confounding variables were fixed to Year 2010-14, Male, Evaluation Score Score AB-A, University, LS in the age interaction model and age lower to 45 for the field interaction model.
  A significant positive percentage change of 10\% for the youngest age group among PIs ($<$ 45) compared to no SNSF funding can be observed for the average number of citations. The remaining changes in citation number are not significant. Then, the effect of funding is largest for SSH researchers (15\% more citations per article as PI compared to unfunded researchers). The effect in LS and STEM is less prominent, +8\% for both. Note that the intervals however all overlap.}}\end{figure}

To better understand those differences in funding effect, we refer to Figure \ref{fig:interactions_articles} for the article counts and Figure \ref{fig:interactions_citations} for the average number of citations per article. Those figures show the predicted article or citation count depending on the funding group (in $t-1$) and the age group or the research field. For all those subgroups, SNSF funding (as PI) in $t-1$ has a positive effect on the outcome. However the size of this effect differs substantially. The youngest age group ($<$45) seems to benefit considerably from the funding in terms of predicted difference between treatment and control researchers in article count, but also in citation per article (the confidence intervals of funded as PI and no funding do not overlap). More senior funded researchers (45-54 and 55-65 years of age) perform similarly well compared to researchers with the same characteristics but no funding. It is noteworthy that for older researchers (65+) the difference between groups is again higher indicating that funding helps to keep productivity up. We obtain very similar results based on post-estimations with interaction effects in the matched samples from the propensity score matching approach, see Figure \ref{fig:combined_matching_interaction}.

For all research areas, SNSF funding has a positive effect on article count and number of citations. STEM researchers however benefit most with a percentage change of 23\% more articles as funded PI compared to no funding; funded (PI) researchers from the LS publish 15\% more articles and the SSH researchers 12\%. This could reflect that in STEM and LS the extent to which research can be successfully conducted is highly funding-dependent, while this is not necessarily the case in the SSH. Yet regarding the number of citations per article, the SSH researchers benefit most (14\% more citations for SSH, 8\% for STEM and 6\% for LS). This suggest that funding may support the quality of research and hence its impact more in the SSH field. Thus, it should be noted that even though SSH researcher publish and are cited less in absolute numbers, we still see a substantial positive effect of SNSF funding on the outcomes. The respective figures for the remaining outcomes can be found in the supplementary material; more specifically Figure \ref{fig:altmetric_interaction} for the altmetric score, Figure \ref{fig:preprint_interaction} for the preprint count and Figure \ref{fig:FCR_interaction} for the FCR.

\hypertarget{conclusions}{%
\section{Conclusions}\label{conclusions}}

Understanding the role played by competitive research funding is crucial for designing research funding policies that best foster knowledge generation and diffusion. By investigating the impact of project funding on scientific output, its relevance and accessibility, this study contributes to research on the effects of research funding at the level of the individual researcher.

Using detailed information - including personal characteristics and the evaluation scores that their submitted projects received by peers - on the population of all project funding applicants at the SNSF during the 2005-2019 period, we estimate the impact of receiving project funding on publication outcomes and their relevance. The strengths of this study are in the detailed information on both researchers and grant proposals. First, the sample consists of both successful as well as unsuccessfully applicants. Therefore researchers who also had a research idea to submit are part of the control group. Second, information on the proposal evaluation scores allows to compare researchers which have submitted project ideas of - on average - comparable quality. 
The estimated treatment effects therefore take into account that all applicants may benefit from the competition for funding through participation effects \citep{Ayoubietal_2019}.

Besides these methodological aspects, a key contribution of this study is that - in addition to articles in scientific journals - it is the first to include preprints. Preprints are an increasingly important means of disseminating research results early and without access restrictions \citep{Bergetal2016,Serghiou2018}. Besides this, we investigate relevance and impact in terms of absolute and relative citation measures. In the analysis of citations that published research receives, it is important to account for field-specific citation patterns. We do so by including the RCR and the FCR as measures for relative research impact in a researcher's own field of study as additional outcome measures. Finally, this is the first study to investigate the link between funding and researchers' altmetrics scores which mirror the attention paid to research outcomes in the wider public \citep{BORNMANN2014,Warrenetal2017,Lazaroiu2017}.

The results show a similar pattern across all estimation methods indicating an effect size of about one additional article in each of the three years following the funding. In addition, we find a similarly sized effect on the number of preprints. The comparison across methods suggests that if accounting for important observable researcher characteristics (e.g. age, field, gender and experience) as well as proposal quality (as reflected in evaluation scores) parametric regression results and non-parametric models lead to similar conclusions with regard to publication outputs. Importantly, a significant effect on the number of citations to articles could be observed indicating that funding does not merely translate into more, but only marginally relevant research. Funded research also appears to reach the general public more than other research as indicated by higher average altmetrics in the group of grant-winners. In terms of the RCR and FCR the results indicate that there might be an effect on the funded researchers' overall visibility in the research community. However, the effects on the RCR are not robust to the estimation method used.   

The funding program analysed in this study is open to all researchers in Switzerland affiliated with institutions eligible to receive SNSF funding. This allows us to study treatment effect heterogeneity over researchers' life cycle and research field. The results suggest here, that funding is particularly important at earlier career stages where PF facilitates research that would not have been pursued without funding.
With regard to treatment effect heterogeneity across fields, we find the highest effect of funding on the article count for STEM researchers and the highest funding effect on citations in SSH.

While the insights on a positive effect of funding on the number of subsequent scientific articles are in line with previous studies, compared to previous results, the effects that we document here are larger. The reason for that may be related to the fact that the SNSF is the main source of research funding in Switzerland we can therefore identify researchers for the control group who really had no other project grant in the period for which they are considered a control. We also observe co-PIs which may in other studies - due to a focus on PIs or lack of information - be assigned to the control group. Both may lead to an under-estimation of funding effects in previous studies. Moreover, by counting all publications of these researchers, we further take not only articles directly related to the project into account, but also that there are learning spillovers and synergies beyond the project that improve a researcher's overall research performance. 

Despite all efforts, this study is not without limitations. First, we do not observe industry funding for research projects which may be important in the engineering sciences \citep{HottenrottThorwarth2011,HottenrottLawson_2017}. 
Moreover, the fact that researchers receive grants repeatedly and may switch between treatment and control group over time, makes a simple difference-in-difference analysis difficult. These factors further complicate the assessment of long-term impact of the research outcomes that we observe.
The methods presented here aim to account for the non-randomness of the funding award and the underlying data structure. While we find that the main results are robust to the estimation method used, the reader should keep in mind that time-varying unobserved factors that affect an individual's publication outcomes such as family or health status, involvement in professional services or administrative roles and duties \citep{Fudickar2016} may be not sufficiently accounted for. Moreover, we do not have detailed information on the involved research teams and individual responsibilities within the projects.  Therefore we do not investigate the role of team characteristics for any outcome effects. In such an analysis, it would be desirable to study whether and how sole-PI and multiple-PI projects differ and which role different PI profiles play for project success. A more detailed analysis of teams would also be interesting in order to differentiate between group and individual effort. Third, we used preprints and altmetrics as output measures which is novel compared to previous research on funding effects. Since we cannot compare our results to previous ones, we encourage future research on the effects of funding on early publishing and science communication more directly. It should be kept in mind that altmetrics may measure popularity in addition to efforts at dissemination as well as the extent to which authors are embedded in a network, but not the quality of individual research outcomes. Probably more than publications in peer-reviewed journals, preprints and altmetrics may be gamed – for example by repeated sharing of own articles or by `Salami slicing' research outcomes into several preprints.  
Finally, it should be noted that we did not investigate several aspects that might be important in impact evaluation in this study. This list includes the role of the funding amount, the degree of novelty of the produced research, as well as treatment effect heterogeneity in terms of individual characteristics other than age. 

\section*{Acknowledgements}
We are grateful to Tobias Phillip for helpful comments on the study design and on previous versions of this manuscript and to Matthias Egger for an additional careful review of the manuscript prior to submission.

\section*{Data availability}
An anonymized and aggregated data set can be found on Zenodo \\ (doi.org/10.5281/zenodo.5011201). In order to anonymise the data we only provide applicants' age as categorical variable.
\section*{Funding}
This work was supported by the SNSF (internal funds).
\section*{Competing interests} None declared.
\newpage
\singlespacing

\pagebreak

\section{Supplementary data description}

Table \ref{tab:output_baseline_RA} shows the same descriptive statistics as in Table \ref{tab:output_baseline} in the main paper, but stratified by research area / field. Table \ref{tab:percentages_found_not_found} shows how the researchers that were found in Dimensions (with one unique ID) differ from the researcher that could not found be unambiguously in Dimensions (no record at all, no unique ID or unmatched other characteristics). Hence, the researchers that were found are slightly older and are more likely to do research in the LS and STEM fields. Regarding the institution type, no large differences are observed.

\begin{table}
\caption{Descriptive statistics on the (yearly) output measures, funding measures and baseline characteristics by main research area.}\label{tab:output_baseline_RA}
\centering \scriptsize
\begin{tabular}{lllrrr}
\toprule
  & Main research area & \% (Mean) & SD & Min & Max\\
\midrule
\addlinespace[0.3em]
\multicolumn{6}{l}{\textbf{Output \& Funding Measures}}\\
\hspace{1em} \# articles & LS & 6.5 & 7.4 & 0 & 141\\
\hspace{1em} & STEM & 5.7 & 8.2 & 0 & 179\\
\hspace{1em} & SSH & 1.5 & 3.7 & 0 & 222\\
\hspace{1em} \# of preprints & LS & 0.1 & 0.7 & 0 & 54\\
\hspace{1em} & STEM & 0.9 & 2.3 & 0 & 41\\
\hspace{1em} & SSH & 0.1 & 0.7 & 0 & 19\\
\hspace{1em} \# of citations & LS & 185.2 & 358.3 & 0 & 7'888\\
\hspace{1em} & STEM & 157.7 & 368.2 & 0 & 6'801\\
\hspace{1em} & SSH & 25.6 &  105.7 & 0 & 3'530\\
\hspace{1em} \# of avg. citations& LS & 5.7 & 5.7 & 0 & 146.2\\
\hspace{1em} & STEM & 4.4 & 4.3 & 0 & 129.4\\
\hspace{1em} & SSH & 1.7 & 3.1 & 0 & 67\\
\hspace{1em} avg. altmetric & LS & 14 & 42.9 & 1 & 2'652.6\\
\hspace{1em} & STEM & 11.5 & 35.2 & 1 & 1'650\\
\hspace{1em} & SSH & 14 & 67.5 & 1 & 4'211\\
\hspace{1em} avg. FCR & LS & 7.2 & 13.1 & 0 & 786.4\\
\hspace{1em} & STEM & 5.7 & 11 & 0 & 786.5\\
\hspace{1em} & SSH & 6.7 & 13.2 & 0 & 522.3\\
\hspace{1em} avg. RCR & LS & 1.8 & 4.2 & 0 & 242.2\\
\hspace{1em} & STEM & 1.1 & 1.9 & 0 & 119.4\\
\hspace{1em} & SSH & 1.3 & 1.9 & 0 & 66.8\\
\rowcolor{gray!11}\hspace{1em} \# of years funded & LS & 4.7 & 5 & 0 & 15\\
\rowcolor{gray!11}
\hspace{1em} & STEM & 5.5 & 5.1 & 0 & 15\\
\rowcolor{gray!11}
\hspace{1em} & SSH & 3.6 & 3.7 & 0 & 15\\
\rowcolor{gray!11}
\hspace{1em} \# of years funded as PI & LS & 3.5 & 4.8 & 0 & 15\\
\rowcolor{gray!11}
\hspace{1em} & STEM & 3.8 & 4.8 & 0 & 15\\
\rowcolor{gray!11}
\hspace{1em} & SSH & 2.4 & 3.3 & 0 & 15\\
\rowcolor{gray!11}
\hspace{1em} \% of treated observations & LS & 0.5 &  &  & \\
\rowcolor{gray!11}
\hspace{1em} & STEM & 0.6 &  &  & \\
\rowcolor{gray!11}
\hspace{1em} & SSH & 0.4 &  &  & \\
\addlinespace[0.3em]
\multicolumn{6}{l}{\textbf{Gender, Age \& Institution Type}}\\
\hspace{1em} Female & LS & 21.7 \% &  &  & \\
\hspace{1em}  & STEM & 14 \% &  &  & \\
\hspace{1em}  & SSH & 34.5\% &  &  & \\
\hspace{1em} Male & LS & 78.3 \% &  &  & \\
\hspace{1em}  & STEM & 86 \% &  &  & \\
\hspace{1em}  & SSH & 65.5\% &  &  & \\
\rowcolor{gray!11}
\hspace{1em} Age & LS & 46.5 & 7.8 &  & \\
\rowcolor{gray!11}
\hspace{1em}  & STEM & 45.7 & 8.5 &  & \\
\rowcolor{gray!11}
\hspace{1em}  & SSH & 47.8 & 8.7 &  & \\
\hspace{1em}Cantonal university & LS & 76.2 \% &  &  & \\
\hspace{1em}  & STEM & 30.5 \% &  &  & \\
\hspace{1em}  & SSH & 65.4 \% &  &  & \\
\hspace{1em} ETH Domain & LS & 10.6 \% &  &  & \\
\hspace{1em}  & STEM & 57.6 \% &  &  & \\
\hspace{1em}  & SSH & 6.5 \% &  &  & \\
\hspace{1em} UAS/UTE/Other & LS & 13.2 \% &  &  & \\
\hspace{1em}  & STEM & 11.9\% &  &  & \\
\hspace{1em}  & SSH & 28.2 \% &  &  & \\
\bottomrule
\end{tabular}
\end{table}

\begin{table}
\caption{Characteristics of the researchers that were found in Dimensions vs. the ones that were not or for which no unique identifier could be found.}\label{tab:percentages_found_not_found}
\centering \footnotesize
\begin{tabular}{lll}
\toprule
  & Found & Not Found\\
  & (8'793) & (2'435) \\
  \hline
\multicolumn{3}{l}{\textbf{Age in 2020}}\\
\hspace{1em}$<$45 & 24.2\% & 16.1\%\\
\hspace{1em}45-54 & 38.2\% & 32.9\%\\
\hspace{1em}55-64 & 27.2\% & 33.6\%\\
\hspace{1em}65+ & 10.4\% & 17.4\%\\
\rowcolor{gray!11}
\multicolumn{3}{l}{\textbf{Main research area}}\\
\rowcolor{gray!11}
\hspace{1em}LS & 38.6\% & 31.6\%\\
\rowcolor{gray!11}
\hspace{1em}STEM & 31.2\% & 25.7\%\\
\rowcolor{gray!11}
\hspace{1em}SSH & 30.1\% & 42.8\%\\
\multicolumn{3}{l}{\textbf{Institution type}}\\
\hspace{1em}Cantonal university & 49.9\% & 52.2\%\\
\hspace{1em}ETH Domain & 20.1\% & 15.9\%\\
\hspace{1em}UAS/UTE/Other & 14.4\% & 16\%\\
\hspace{1em}Unclassified & 15.6\% & 15.8\%\\
\bottomrule
\end{tabular}
\end{table}

\section{Supplementary Analyses}\label{supp-analyses}
\subsection{Linear Feedback Models}
As an alternative to estimating random intercepts to account for the heterogeneity of the researchers, we can model past performance as a main driver of current performance. These models are referred to as linear feedback models (LFM). That is, we employ pre-sample information of the researcher as a proxy for unobservable characteristics, such as a researcher's ability or writing talent which impact research output in the (later) sample period. Past performance thereby captures differences in unobserved researcher characteristics such as ability, skills, and ambition \citep{Blundell,HottenrottThorwarth2011}. 
Thus, if interested in counts of articles, the average number of yearly articles in the 5-year window before the start of the observation period can be used to account for unobserved time-invariant differences between researchers. 
In such a LFM, the researcher-specific \emph{stock} of the dependent variable, \(I_{i}\), is used as a proxy for \(v_i\). For the count variables, the model is defined as follows
\[\log \mbox{E} (P_{it} \ | \ \mbox{data}) =  \phi \ [F_{it-1} + X_{it} + T_t +  I_{i}] \ .\]
Note that for the LFMs, robust standard errors have to be calculated due to over-dispersion in the dependent variables\footnote{To do so, we use the \texttt{pglm()} \citep{Croissant} from a package dedicated to panel data econometrics in \texttt{R}.}.
The detailed results can be found in Tables \ref{tab:model_lfm_ols} and \ref{tab:model_lfm_irr_count}. For the count outcomes (peer-reviewed publications and preprints) as well as for the citation outcome (average yearly number of citations per publication) and the altmetric the results are very similar and lead to the same conclusions as the mixed models presented in the main paper with respect to SNSF funding as PI. For the FCR, however, the effect sizes decline in response to adding pre-sample information, from 2\% to about 1.1\%. For the RCR, evidence for an effect of SNSF funding is again insignificant. 

\begin{table}
\caption{\textbf{Incidence rate ratios} (IRR) from the multivariate \textbf{\textit{Poisson Linear Feedback}} models for the count outcomes}\label{tab:model_lfm_irr_count}
\centering
\resizebox{1.0\linewidth}{!}{
\begin{tabular}[t]{lrllrll}
\toprule
\multicolumn{1}{c}{ } & \multicolumn{3}{c}{1. articles} & \multicolumn{3}{c}{2. preprints} \\
\multicolumn{1}{c}{ } & \multicolumn{3}{c}{(72'738 obs.)} & \multicolumn{3}{c}{(61'726 obs.)} \\
\cmidrule(l{3pt}r{3pt}){2-4} \cmidrule(l{3pt}r{3pt}){5-7}
  & IRR & (95\%-CI) & p-val. & IRR  & (95\%-CI)  & p-val. \\
\rowcolor{gray!6}
\midrule
Funding as PI (t-1) & 1.19 & (1.17; 1.2) & $<$ 0.001 & 1.14 & (1.09; 1.19) & $<$ 0.001\\
\rowcolor{gray!6}
Funding as Co-PI (t-1) & 1.11 & (1.09; 1.12) &  & 1.09 & (1.04; 1.15) & \\
Eval. score BC-B & 1.00 & (0.98; 1.01) & 0.8772 & 0.93 & (0.9; 0.97) & $<$ 0.001\\
Eval. score C-D & 1.00 & (0.98; 1.01) &  & 0.74 & (0.68; 0.8) & \\
\rowcolor{gray!6}
Male researcher & 1.37 & (1.3; 1.45) & $<$ 0.001 & 1.77 & (1.54; 2.03) & $<$ 0.001\\
Age (decades) & 1.08 & (1.06; 1.1) & $<$ 0.001 & 1.35 & (1.28; 1.43) & $<$ 0.001\\
\rowcolor{gray!6}
ETH Domain & 0.98 & (0.92; 1.05) & $<$ 0.001 & 1.35 & (1.18; 1.56) & $<$ 0.001\\
\rowcolor{gray!6}
UAS/UTE/Other & 0.66 & (0.62; 0.71) &  & 0.37 & (0.31; 0.44) & \\
\rowcolor{gray!6}
Unclassified & 1.05 & (0.98; 1.12) &  & 1.38 & (1.19; 1.61) & \\
Research Area MINT & 0.82 & (0.77; 0.87) & $<$ 0.001 & 4.04 & (3.55; 4.58) & $<$ 0.001\\
Research Area SSH & 0.33 & (0.31; 0.35) &  & 0.99 & (0.85; 1.14) & \\
\rowcolor{gray!6}
Year 2010-14 & 1.14 & (1.12; 1.15) & $<$ 0.001 &  &  & \\
\rowcolor{gray!6}
Year 2015-19 & 1.16 & (1.14; 1.18) &  &  &  & \\
\rowcolor{gray!6}
Year 2015-19 &  &  &  & 1.52 & (1.46; 1.58) & $<$ 0.001\\
Pre-sample articles & 7.07 & (6.56; 7.61) & $<$0.001 &  &  & \\
Pre-sample preprints &  &  &  & 11.59 & (9.22; 14.57) & $<$ 0.001\\
\bottomrule
\end{tabular}}
   \begin{threeparttable}
      \begin{tablenotes}
        \small  \item Notes: IRR stands for Incidence Rate Ratio. 95\% confidence intervals (CI) provided. The p-values refer to the results of likelihood ratio tests for each variable to be present in the model. For the institution type University serves as reference category, for the fields it is Life Sciences. For the evaluation score the class AB-A serves as reference category and for year it is the period 06-09 for model 1 and 10-14 for model 2.
        \end{tablenotes} 
      \end{threeparttable}
\end{table}

\begin{landscape}
\begin{table}
\caption{Coefficients and percentage changes from  
        \textbf{\textit{linear feedback models}} for the continuous outcomes.}
        \label{tab:model_lfm_ols}
\centering
\resizebox{1.0\linewidth}{!}{
\begin{tabular}[t]{lrllrllrllrll}
\toprule
\multicolumn{1}{c}{ } & \multicolumn{3}{c}{3. Av. cit. per article} & \multicolumn{3}{c}{4. altmetric} & \multicolumn{3}{c}{5. RCR} & \multicolumn{3}{c}{6. FCR} \\
\multicolumn{1}{c}{ } & \multicolumn{3}{c}{(72'738 obs.)} & \multicolumn{3}{c}{(37'273 obs.)} & \multicolumn{3}{c}{(23'350 obs.)} & \multicolumn{3}{c}{(41'874 obs.)} \\
\cmidrule(l{3pt}r{3pt}){2-4} \cmidrule(l{3pt}r{3pt}){5-7} \cmidrule(l{3pt}r{3pt}){8-10} \cmidrule(l{3pt}r{3pt}){11-13}
  & Coef & (95\%-CI) & p-val. & \% Change & (95\%-CI)  & p-val.  & \% Change  & (95\%-CI)   & p-val.   & \% Change   & (95\%-CI)    & p-val.   \\
\rowcolor{gray!6}
\midrule
Funding as PI (t-1) & 0.33 & (0.27; 0.4) & $<$ 0.001 & 5.3 & (1.9; 8.9) & 0.0079 & 0.4 & (-1.2; 1.9) & 0.8788 & 1.1 & (-0.7; 2.9) & 0.0917\\
\rowcolor{gray!6}
Funding as Co-PI (t-1) & 0.23 & (0.16; 0.31) &  & 2.5 & (-1.5; 6.7) &  & 0.0 & (-1.9; 1.9) &  & 2.3 & (0.2; 4.6) & \\
Eval. score BC-B & -0.04 & (-0.12; 0.04) & $<$ 0.001 & -10.8 & (-14.3; -7.3) & $<$ 0.001 & -3.7 & (-5.8; -1.6) & $<$ 0.001 & -2.3 & (-4.3; -0.3) & $<$ 0.001\\
Eval. score C-D & -0.21 & (-0.32; -0.11) &  & -19.7 & (-23.7; -15.6) &  & -8.2 & (-10.5; -5.9) &  & -5.7 & (-8.2; -3.2) & \\
\rowcolor{gray!6}
Male researcher & -0.01 & (-0.22; 0.2) & 0.9148 & 8.5 & (3.2; 14) & 0.0014 & 1.0 & (-1.6; 3.5) & 0.4584 & 3.6 & (0.5; 6.8) & 0.0223\\
Age (decades) & 0.54 & (0.46; 0.62) & $<$ 0.001 & -3.1 & (-5.2; -1) & 0.0040 & -3.5 & (-4.6; -2.3) & $<$ 0.001 & -7.2 & (-8.4; -6) & $<$ 0.001\\
\rowcolor{gray!6}
ETH Domain  & 0.60 & (0.35; 0.86) & $<$ 0.001 & 12.8 & (6.8; 19.1) & $<$ 0.001 & 7.9 & (4.2; 11.8) & $<$ 0.001 & 3.4 & (0.2; 6.7) & $<$ 0.001\\
\rowcolor{gray!6}
UAS/UTE/Other & -0.58 & (-0.84; -0.33) &  & -1.4 & (-7.7; 5.4) &  & -4.3 & (-7.4; -1.1) &  & -5.9 & (-9.6; -2.1) & \\
\rowcolor{gray!6}
Unclassified & -0.13 & (-0.38; 0.13) &  & 8.8 & (2.5; 15.4) &  & 1.5 & (-1.6; 4.7) &  & 0.7 & (-2.8; 4.3) & \\
MINT  & -1.60 & (-1.82; -1.37) & $<$ 0.001 & -32.3 & (-35.6; -28.9) & $<$ 0.001 &  &  &  & -16.1 & (-18.5; -13.7) & $<$ 0.001\\
SSH & -3.04 & (-3.27; -2.81) &  & -22.3 & (-26.5; -17.9) &  &  &  &  & -11.2 & (-14.2; -8) & \\
\rowcolor{gray!6}
Year 2010-14 & 0.78 & (0.71; 0.85) & $<$ 0.001 &  &  &  & 0.6 & (-0.9; 2.1) & 0.757 & -8.3 & (-10; -6.7) & $<$ 0.001\\
\rowcolor{gray!6}
Year 2015-19 & 1.29 & (1.2; 1.39) &  &  &  &  & 0.5 & (-1.2; 2.2) &  & -22.3 & (-23.8; -20.8) & \\
\rowcolor{gray!6}
Year 2015-19  &  &  &  & 50.7 & (47; 54.5) & $<$ 0.001 &  &  &  &  &  & \\
Pre-sample citation count & 2.92 & (2.67; 3.17) & $<$ 0.001 &  &  &  &  &  &  &  &  & \\
Pre-sample altmetric &  &  &  & 99.7 & (85.9; 114.5) & $<$ 0.001 &  &  &  &  &  & \\
Pre-sample RCR &  &  &  &  &  &  & 14.7 & (10.5; 19) & $<$ 0.001 &  &  & \\
Pre-sample FCR &  &  &  &  &  &  &  &  &  & 37.7 & (35.4; 39.9) & $<$ 0.001\\
\bottomrule
\end{tabular}}
   \begin{threeparttable}
      \begin{tablenotes}
        \small  \item Notes: The coefficients of the models
        on the logarithmic scales of the outcomes are converted to percentage changes, while the coefficient for the average citation per publications stays a unit change. The quantities are provided together with a 95\% confidence intervals (CI). The p-values refer to the results of likelihood ratio tests for each variable to be present in the model. For the institution type University serves as reference category, for the fields it is Life Sciences. For the evaluation score the class AB-A serves as reference category and for year it is the period 06-09 for model 3, 5, 6 and 10-14 for model 4.
        \end{tablenotes} 
      \end{threeparttable}
\end{table}
\end{landscape}

\subsection{Additional material for propensity score analysis}

Table \ref{tab:drawing} shows the count statistics for the drawing with replacement in the propensity score matching analysis. Hence, across the different matched samples less than 10\% of control researcher-year observations are used only once and about 60\% up to 25 times. About 10\% of control group researchers are used very frequently, i.e. about 200 times. In the FCR sample this number is slightly higher with about 15\% indicating that here some researchers serves as 'twin' for many treated researchers. The patterns for the preprint and the RCR sub-samples are very similar to the ones for the  samples presented in the Table.

\begin{figure}
{\centering \includegraphics[width=.9\linewidth]{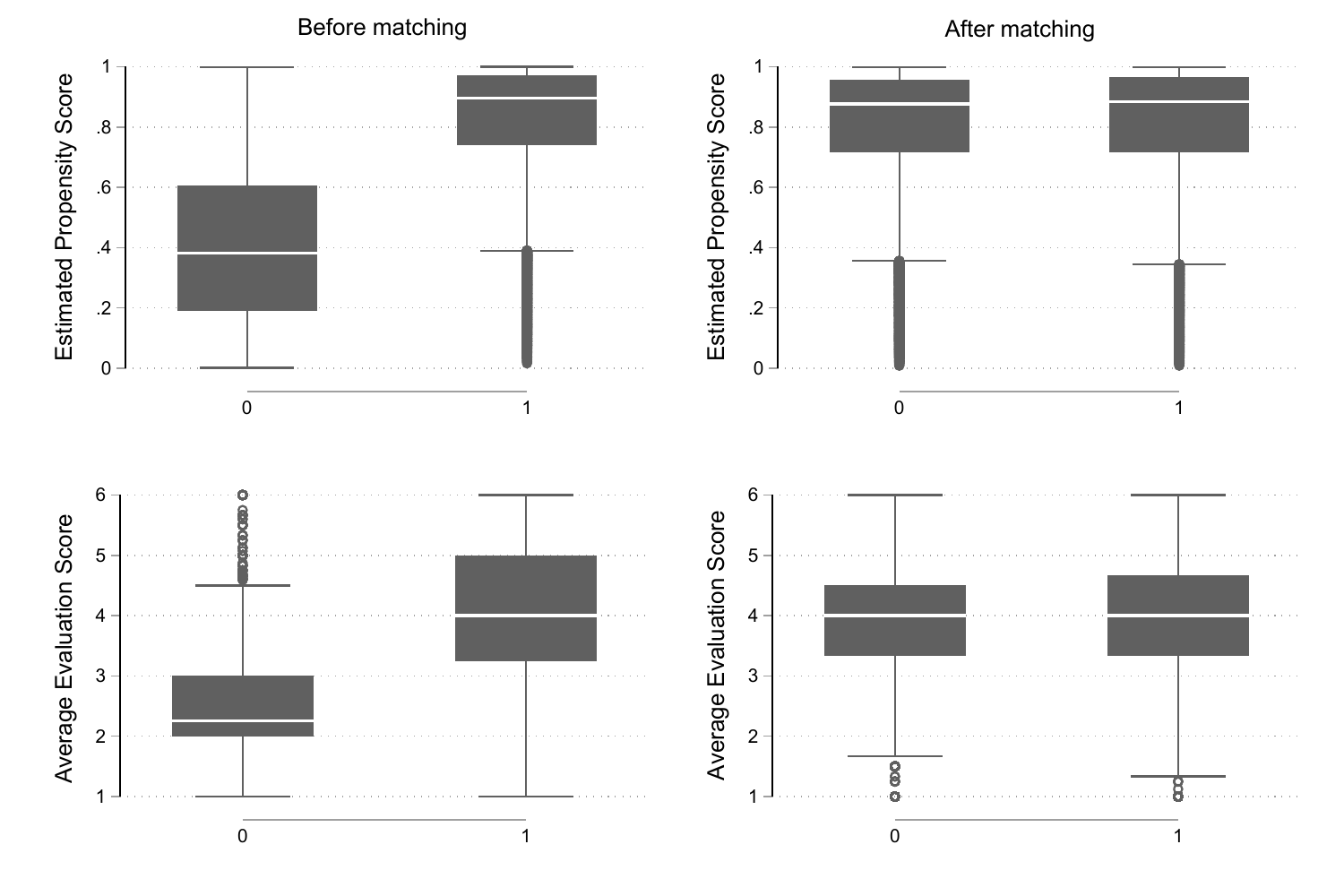}}
\caption{Balancing of the propensity score and the evaluation score (n=63'680)}.\label{fig:matchingsuccess}
\end{figure}

Figure \ref{fig:matchingsuccess} shows the comparison of the distribution of the propensity score and the evaluation score before and after matching and therefore shows that the nearest neighbor matching procedure was successful in balancing the sample in terms of the grant likelihood and - importantly - also the average scores.

\begin{table}
  \centering \scriptsize
  \caption{Recurrent matching (matches per control group individual)}
        \label{tab:drawing}
           \begin{tabular}{lrrr}
    \toprule
        Times matched  & \multicolumn{1}{l}{Freq.} & \multicolumn{1}{l}{Percent} & \multicolumn{1}{l}{Cum.} \\
         \midrule
    Articles sample &       &       &  \\
    \midrule
    1                    & 3'050 & 6.95  & 6.95 \\
    1 $<$ x $\leq$ 10    & 13'073 & 32.57 & 39.52 \\
    10 $<$ x $\leq$ 25   & 10'908 & 21.98 & 61.50 \\
    25 $<$ x $\leq$ 50   & 6'105  & 13.91 & 75.41 \\
    50 $<$ x $\leq$ 100  & 6'116  & 13.93 & 89.34 \\
    100 $<$ x $\leq$ 213 & 4'684  & 10.66   & 100 \\
    \midrule
    Total & 43'936 & 100   &  \\
     \midrule
    Altmetrics sample &       &       &  \\
    \midrule
    1 & 1'274 & 5.71  & 5.71 \\
    1 $<$ x $\leq$ 10 & 6'375 & 28.50 & 34.21 \\
    10 $<$ x $\leq$ 25 & 4'513 & 20.17 & 54.38 \\
    25 $<$ x $\leq$ 50 & 4'470  & 20.02 & 74.40 \\
    50 $<$ x $\leq$ 100 & 3'201  & 14.30 & 88.70 \\
    100 $<$ x $\leq$ 164 & 2'529  & 11.30   & 100 \\
    \midrule
    Total & 22'362  & 100   &  \\
       \midrule
    FCR sample &       &       &  \\
    \midrule
    1 & 1'917 & 6.62  & 6.62 \\
    1 $<$ x $\leq$ 10 & 9'075 & 31.36  & 37.99 \\
    10 $<$ x $\leq$ 25 & 5'983  & 20.68 & 58.66 \\
    25 $<$ x $\leq$ 50 & 3'827  & 13.23  & 71.89 \\
    50 $<$ x $\leq$ 100 & 3'724  & 12.87 & 84.76 \\
    100 $<$ x $\leq$ 191 & 4'410  & 15.24 & 100 \\
    \midrule
    Total & 28'936 & 100   &  \\
    \bottomrule
    \end{tabular}%
\end{table}%

\section{Additional Figures}
The following section includes additional Figures to complete the data description as well as the results sections of the main paper.

\subsection{Descriptive Statistics}
\begin{figure}
    \centering
     \includegraphics[width=1\linewidth]{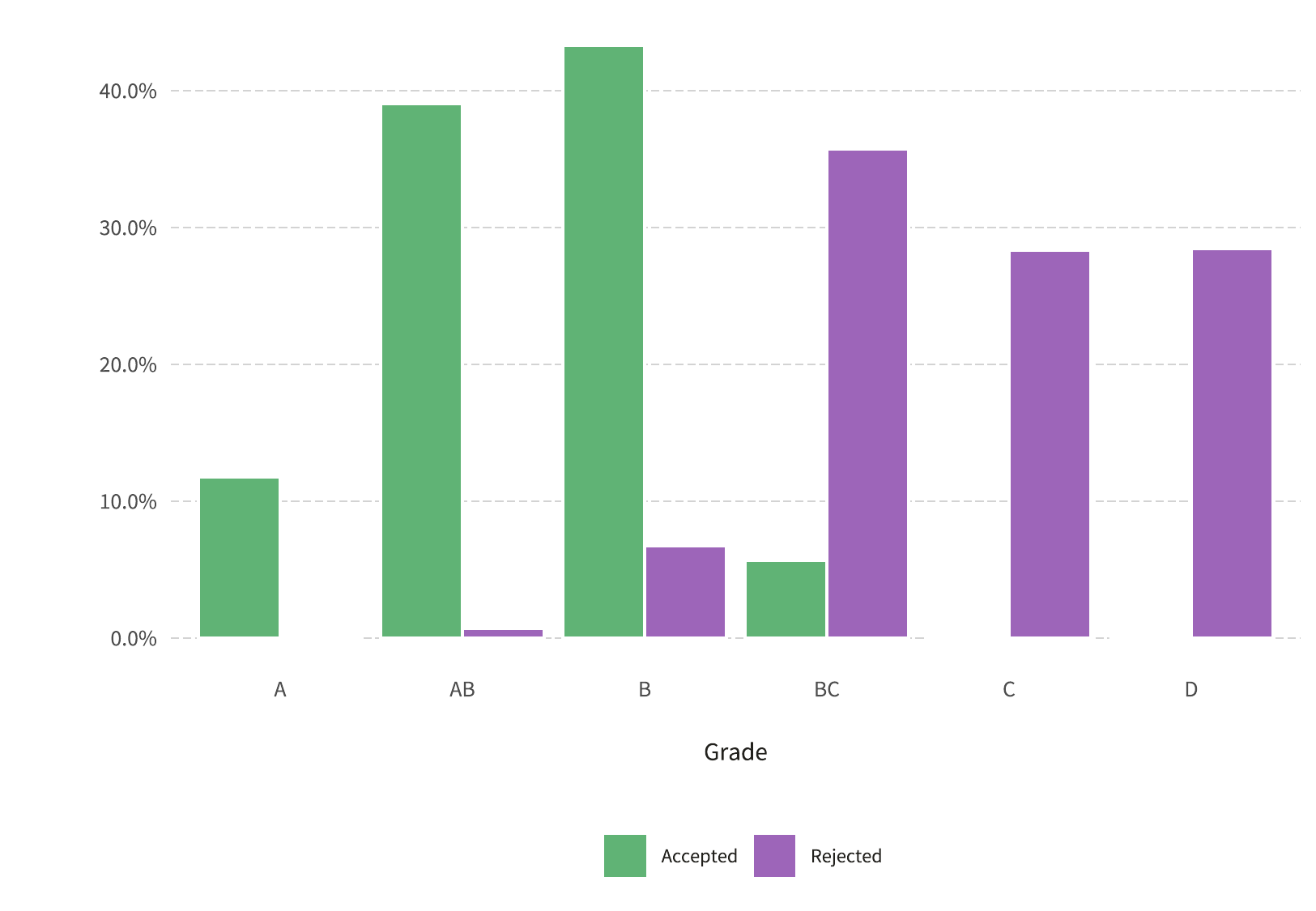}
    \caption{Distribution of the grades among the rejected (purple) and accepted (green) proposals.}
    \label{fig:grade_distribution}
\end{figure}

\begin{figure}
    \centering
     \includegraphics[width=1\linewidth]{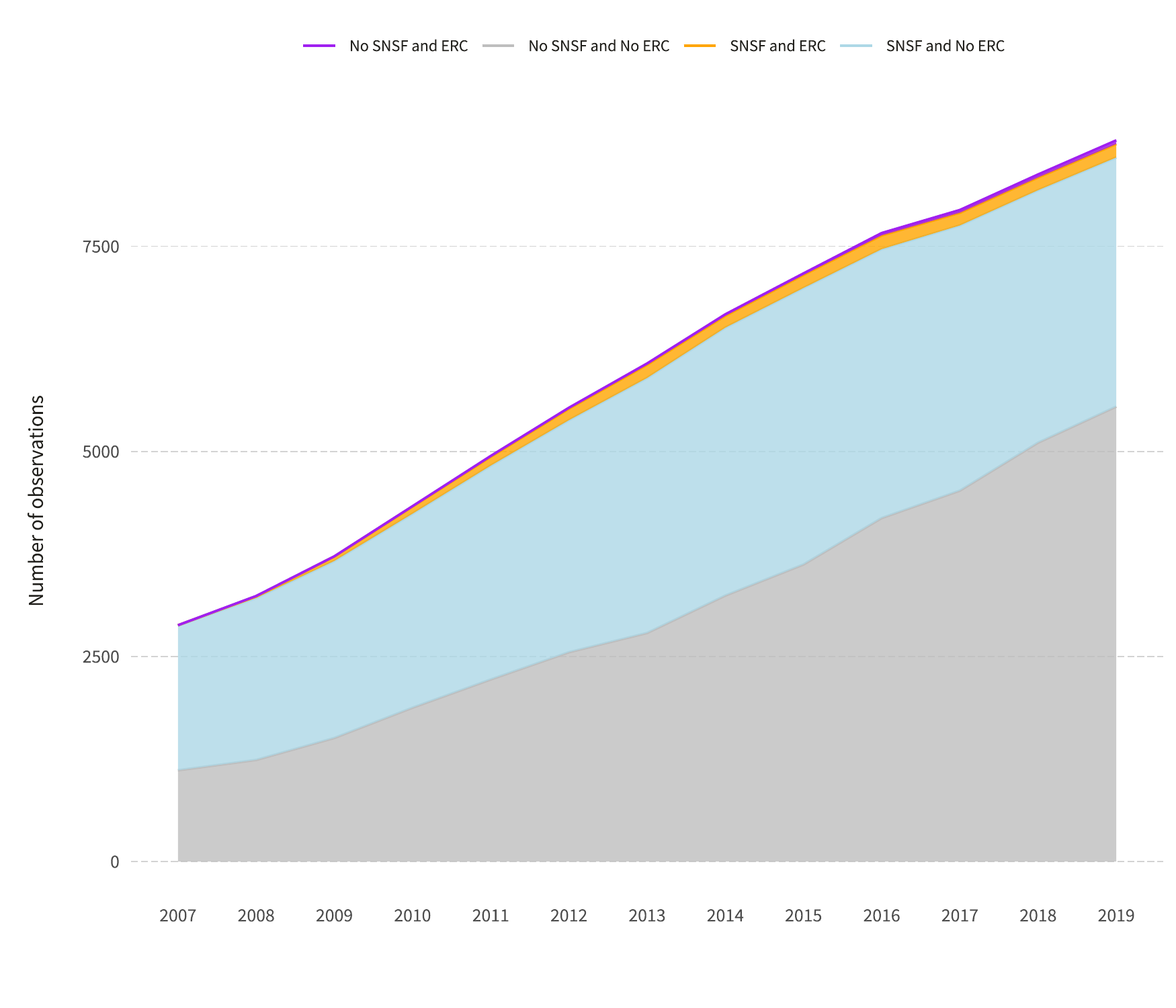}
    \caption{Evolution of the count of observations in different funding groups.}
    \label{fig:erc_evolution}
\end{figure}

Figure \ref{fig:grade_distribution} represents the distribution of the grades among rejected and accepted projects. Figure \ref{fig:erc_evolution} shows the count of observations in the different funding, meaning researchers who are funded by the ERC and the SNSF simultaneously, only by the ERC or the SNSF, or by neither of those funders.

\subsection{Interaction effects}
\begin{figure}
\centering
     \includegraphics[width=1\linewidth]{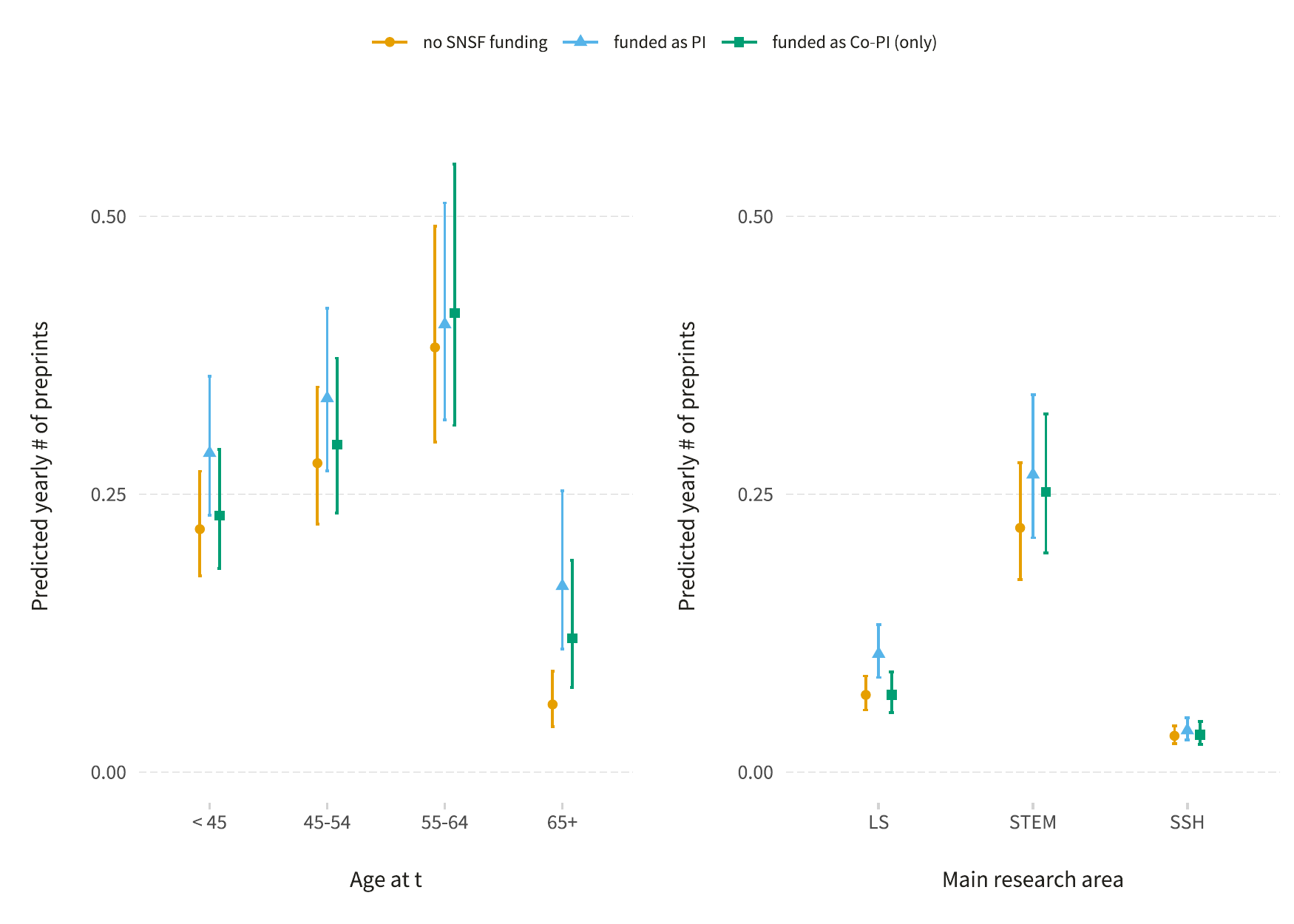}
     \caption{Predicted number of preprints per researcher depending on whether she was treated (as PI/Co-PI) or not (no funding) and her age group (left) or the research field (right). Younger and older generations benefit more from SNSF funding with respect to preprints. To predict the annual number of preprints the baseline confounding variables were fixed to Year 2015-19, Male, Evaluation Score Score AB-A, ETH Domain, MINT for the age interaction model and age smaller than 45 for the field interaction model.}
     \label{fig:preprint_interaction}
\end{figure}

\begin{figure}
\centering
     \includegraphics[width=1\linewidth]{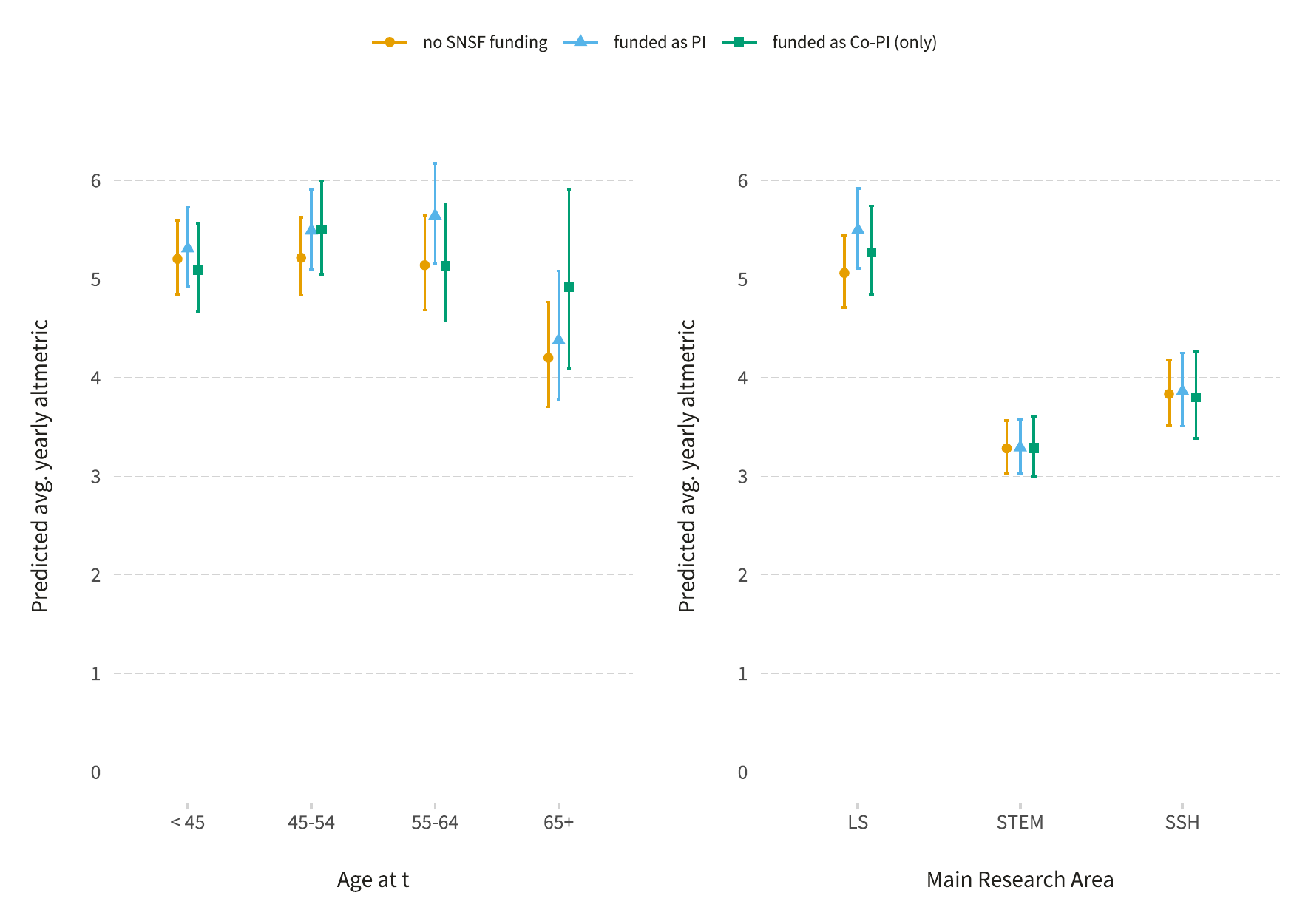}
\caption{Predicted average altmetric score per researcher depending on whether she was treated (as PI/Co-PI) or not (no funding) and her age group (left) or the research field (right). No conclusive results on interaction between SNSF-funding  and altmetrics. To predict the researchers annual average altmetric score the confounding variables were fixed to Year 2010-14, Male, Evaluation Score Score AB-A, University, LS for the age group interaction model and age smaller than 45 for the research field interaction model.}
\label{fig:altmetric_interaction}
\end{figure}

\begin{figure}
\centering
     \includegraphics[width=1\linewidth]{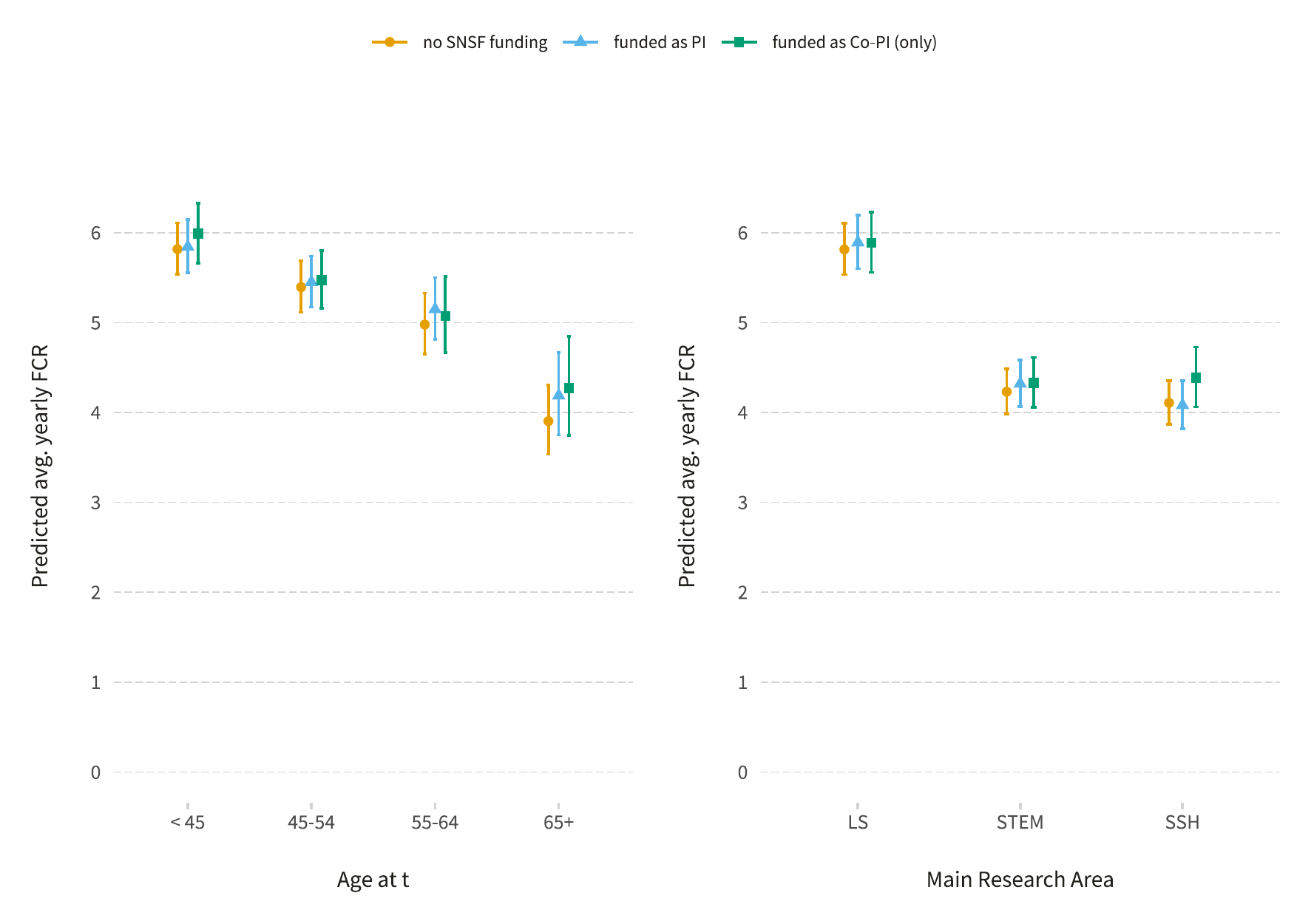}
\caption{Predicted average FCR depending on whether the researcher was treated (as PI/ Co-PI) or not (no funding), and her age group (left) or the research field (right). No conclusive results on interaction between SNSF-funding and FCR. To predict the researchers annual average FCR the confounding variables were fixed to Year 2010-14, Male, Evaluation Score Score AB-A, University, LS for the age group interaction model and age smaller than 45 for the research field interaction model.}
\label{fig:FCR_interaction}
\end{figure}

\begin{figure}
{\centering \includegraphics[width=1.2\linewidth]
{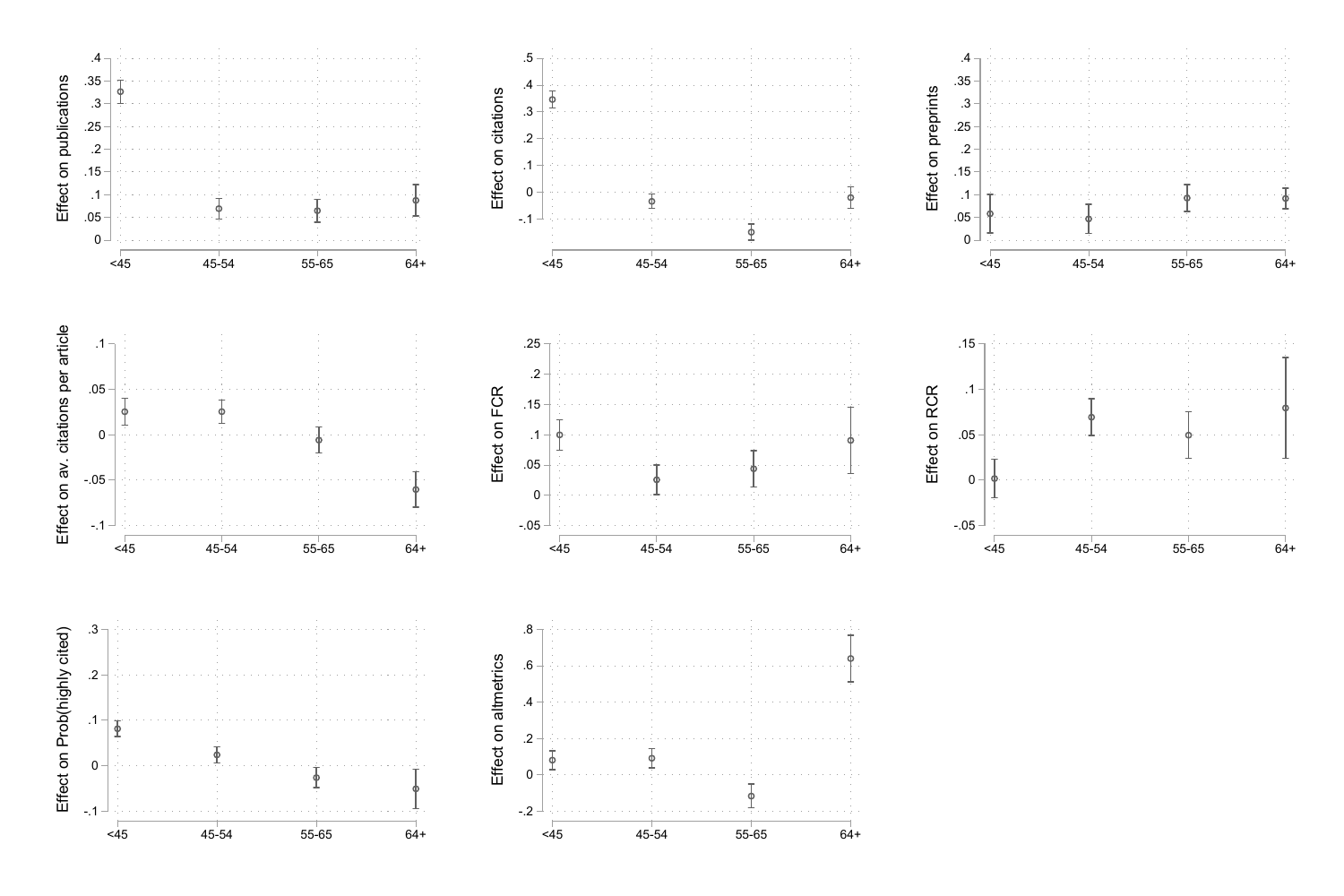}}
\caption{Estimated average marginal effects of funding based on random effects model after matching by age group (outcome variables measured in $t+1$).} \label{fig:combined_matching_interaction}
\end{figure}

The differences in the funding effect with respect to the age of the researcher and the research field were studies also for the yearly number of preprints (Figure \ref{fig:preprint_interaction}), the altemtric score (Figure \ref{fig:FCR_interaction}) and the FCR (Figure \ref{fig:altmetric_interaction}). Finally, Figure \ref{fig:combined_matching_interaction} represents the same results using the matching exercise.

\end{document}